\documentclass[preprint,review,12pt]{elsarticle}

\usepackage{amssymb}

\usepackage{graphicx}
\usepackage{amsmath}
\usepackage{amssymb}
\usepackage{psfrag}
\usepackage{setspace}
\usepackage{wrapfig}
\usepackage{sidecap}

\usepackage{caption}
\usepackage{subcaption}

\newcommand{\bs}{\boldsymbol}
\def\RR{ \mathbb R}

\usepackage{epsf}
\newcommand{\ee}{\end{equation}}
\newcommand{\be}{\begin{equation}}
\newcommand{\ec}{\end{center}}
\newcommand{\bc}{\begin{center}}
\newcommand{\eea}{\end{eqnarray}}
\newcommand{\bea}{\begin{eqnarray}}
\newcommand{\bd}{\begin{description}}
\newcommand{\ed}{\end{description}}
\newcommand{\bi}{\begin{itemize}}
\newcommand{\ei}{\end{itemize}}
\newcommand{\pa}{\partial}

\newcommand{\bt}{\boldsymbol{\theta}}
\newcommand{\bz}{\boldsymbol{z}}
\newcommand{\by}{\boldsymbol{y}}
\newcommand{\bez}{\boldsymbol{\eta}_z}
\newcommand{\bet}{\boldsymbol{\eta}_{\theta}}
\newcommand{\bmz}{\boldsymbol{\mu}_z}
\newcommand{\bmt}{\boldsymbol{\mu}_{\theta}}
\newcommand{\dt}{d_{\theta}}
\newcommand{\bgt}{\bs{G}_{\theta}}
\newcommand{\bgz}{\bs{G}_{z}}

\newcommand{\refeq}[1]{Equation (\ref{#1})}

\usepackage[usenames,dvipsnames]{color}

\journal{-}

\begin{document}

\begin{frontmatter}



\title{Variational Bayesian strategies for high-dimensional,
stochastic design problems}


\author[rvt]{P.S. Koutsourelakis\corref{cor1}}
\ead{p.s.koutsourelakis@tum.de}
\cortext[cor1]{Corresponding Author. Tel: +49-89-289-16690}
\address[rvt]{Professur  f\"ur Kontinuumsmechanik, Technische Universit\"at M\"unchen, Boltzmannstrasse 15, 85747 Garching (b. M\"unchen), Germany}
\ead[url]{http://www.contmech.mw.tum.de}

\begin{abstract}
This paper is concerned with a lesser-studied problem in the context of model-based, uncertainty quantification (UQ), that of optimization/design/control under uncertainty. The solution of such problems is hindered not only by the usual difficulties encountered in UQ tasks (e.g. the high computational cost of each forward simulation, the large number of random variables) but also by the need to solve a nonlinear optimization problem involving large numbers of design variables and potentially constraints.
We propose a framework that is suitable  for  a large class of such problems  and  is based on the idea of recasting them as probabilistic inference tasks. To that end, we propose a Variational Bayesian (VB) formulation and an iterative VB-Expectation-Maximization scheme that is also capable of identifying a low-dimensional set of directions in the design space,
 along  which, the objective exhibits the largest sensitivity. 
 We demonstrate the validity of  the proposed approach in the context of two numerical examples involving $\mathcal{O}(10^3)$ random and design variables. In all cases considered the cost of the computations in terms of calls to the forward model was of the order $\mathcal{O}(10^2)$. The accuracy of the approximations provided is assessed by appropriate information-theoretic metrics.\footnote{
This paper is based on the homonymous talk given during the international symposium on "Big Data and Predictive Computational Modeling" that took place in 18-21 May 2015 at TUM-IAS, Munich Germany.}

\end{abstract}


\end{frontmatter}


\section{Introduction-Motivation}
\label{sec:intro}

With the increased computational capabilities afforded by the  utilization of peta- and exa-scale computing resources throughout engineering and the physical sciences, the issue of confidence in simulation results has come at the center of current research. The objective of obtaining a nominal  computational representation of a physical process is being replaced by the new paradigm of {\em predictive simulations} where the analysis delivers a quantification of uncertainty due to randomness  in parameters, data or models. Decisions that are based on high-fidelity computational simulations 
due to their potential economic or societal impact cannot be accepted without quantitative information on the confidence in the computed result.

The field of model-based, uncertainty quantification has seen marked advances in recent years. Naturally, the majority of the efforts have been directed towards {\em forward} uncertainty propagation
 i.e. the computation of output statistics given input uncertainties. While several important challenges still remain unanswered, the ultimate objective of the analysis of physical processes  and engineering systems is to enable their control and optimization with respect to design objectives. 
 Problems of optimization in the presence of uncertainty have attracted much less attention.
 On one hand, this is because they encompass all the difficulties encountered in uncertainty propagation. First and foremost the complexity of the forward problem and the increased computational expense associated which each call to the forward solver. It is generally the number of such forward solves that determines the overall computational cost. Secondly, the high-dimensionality of the vector of random variables. Especially in cases where spatiotemporal discretizations of random processes and fields are necessary, one must frequently deal with thousands of random variables. 
 Furthermore, in stochastic optimization problems,  there is  the additional need to solve  a demanding, nonlinear optimization problem which might itself involve thousands of design variables as well as equality/inequality constraints,
 
Significant advances have been achieved in deterministic optimization and control of complex systems particularly with the development of adjoint-based techniques \cite{biegler_large-scale_2003,akcelik_adjoint_2005,hinze_optimization_2009} as well as by making use of reduced-order modeling techniques \cite{dihlmann_certified_2014,yongjin_zhang_accelerating_2015}. Nevertheless their direct application in the stochastic counterparts of these problems would be infeasible or impractical as the integration with respect to uncertainties poses an insurmountable task. 

While decision-making under uncertainty was pioneered in the 1950s \cite{dantzig_linear_1955}, applications to large-scale physical models are scarce due to the inherent computational difficulties.
Advances in stochastic/robust control and optimization \cite{tsompanakis_structural_2008,lagaros_multi-objective_2005,beyer_robust_2007} or reliability-based design optimization  \cite{papadrakakis_design_2005,papadrakakis_reliability-based_2002} are generally applicable to small systems or rely on specific system structure. Techniques using surrogate models and response surfaces \cite{sankaran_method_2010} or generalized Polynomial Chaos expansions \cite{sankaran_stochastic_2009} might  fail to provide good approximations if the number of uncertainties is large, irreducible or non-Gaussian. Furthermore, there is a difficulty in quantifying the error introduced due to the discrepancy between the surrogate and reference model. A critical problem in that respect is the ability to deal with noisy evaluations of the objective functions, its gradient and higher-order derivatives. 
 
 The stochastic optimization framework advocated in the present paper is motivated by the following desiderata:
 \bi
 \item The ability to seamlessly utilize deterministic (legacy) simulators and deterministic optimization components such as a first and second order parametric derivatives of model outputs.

\item The ability to deal with high-dimensional vectors of random and design variables.

\item Least possible number of forward solutions 
\item The ability to quantify the robustness of the identified optimum and provide information on the design features that exhibit the largest sensitivity.
\item The ability to  utilize even highly-approximate, reduced-order models or surrogates  in order to expedite the solution process. 
\ei

 The objective functions considered in this paper can be written in a general form as:
 \be
 V(\bz)=\int U(\bt,\bz) ~p_{\theta}(\bt)~d\bt
 \label{eq:main}
 \ee
 where $\bt \in \RR^{d_{\theta}}$ denotes the vector of random variables with a probability density function $p_{\theta}(\bt)$ and $\bz \in \RR^{d_z}$ denotes the vector of design variables.
 The  function $U(\bt,\bz)$ depends on the {\em output} of the mathematical  model and in turn,   implicitly depends on random and design variables. Each evaluation of $U(\bt,\bz)$ implies a forward model solution which is assumed {\em expensive as in most challenging applications}.
 Naturally the optimization problem can be augmented with constraints with regards to the design variables as it will be demonstrated in the stochastic topology optimization problem that will be considered in the last  section.
 We adopt the term {\em utility function} (opposite  of a loss function) for $U(\bt,\bz)$ and expected utility for $V(\bz)$  and,   without loss of generality,  pose the corresponding problem as one of {\em maximization}.
 
 The formulation above is quite general and can be readily adapted to cases of practical interest. For example if $U(\bt, \bz)=\bs{1}_{\mathcal{A}}(\bt, \bz)$ is the indicator function of an event $\mathcal{A}$ of interest (e.g. failure, or exceedance of a response threshold) then maximizing $V(\bs{z})$ in \refeq{eq:main} is equivalent to the maximization of the probability associated with the event $\mathcal{A}$ (similarly one can minimize the probability of event  $\mathcal{A}$ by employing the indicator function of the complementary even $\mathcal{A}^c$ in  place of $U$ in \refeq{eq:main}). 
 The case that would be of principal concern in this paper involves utility functions of the following form \footnote{As it will become apparent in the subsequent derivations, the exponent in \refeq{eq:util} is used in order to simplify the presentation and several other options to the same effect are possible.}:
 \be
 U(\bt,\bz) =\exp \{ - \frac{1}{2}\parallel \bs{Q}^{1/2} (\bs{u}_{target}-\bs{u}(\bt, \bz))\parallel^2  \}
 \label{eq:util}
 \ee
 where $\bs{u}(\bt, \bz) \in \RR^n$ denotes an output vector of interest (i.e. displacements, velocities, temperature etc), $\bs{u}_{target} \in  \RR^n$ a target/desired response and  $\bs{Q}$  a positive definite matrix of choice (in the current examples $\bs{Q}=\tau_Q \bs{I}_n$). Maximizing the corresponding expected utility implies finding $\bs{z}$ for which the response quantities of interest are, {\em on average}, as close (in the norm defined by $\bs{Q}$) to the target values $\bs{u}_{target}$.
 Similar objective functions have been employed by \cite{hyun_designing_2001} to identify random composites with target effective/homogenized properties and  in \cite{capiez-lernout_robust_2008} in the context of computational mechanics.  
In addition,  related  stochastic design/control objectives have been proposed in \cite{zabaras_scalable_2008} and \cite{seshadri_aggressive_????}.
 
 The obvious strategy for maximizing the expected utility in \refeq{eq:main} is stochastic approximations such as noisy gradient ascent which, in its simplest form,  iterates as follows:
 \be
 \bz^{(t+1)}=\bz^{(t)}+\eta_t \hat{\bs{g}}_t
 \ee
 where $\hat{\bs{g}}_t$ is a noisy (unbiased) estimator of the gradient:
 \be
 \nabla_{\bz} V(\bz^{(t)}) = \int \frac{\pa U(\bt, \bz^{(t)}) }{\pa \bz}~p_{\theta}(\bt)~d\bt
 \ee
 and $\eta_t$ a sequence of learning rates that satisfy $\sum_{t=0}^{\infty} \eta_t=+\infty$ , $ \sum_{t=0}^{\infty} \eta_t^2< +\infty$ \cite{robbins_stochastic_1951,kiefer_stochastic_1952}. While convergence to a (local) maximum is assured under fairly weak conditions  \cite{spall_introduction_2003,kushner_stochastic_2003} even when a single sample of $\bt$  from $p_{\theta}(\bt)$ is used in the context of a basic Monte Carlo estimate of $\hat{\bs{g}}_t$, the convergence rate can be slow requiring an exuberant number of forward calls to evaluate $U$ and/or $\frac{\pa U }{\pa \bz}$. 
 
 An alternative perspective to the problem was proposed in \cite{muller_simulation_1998} where  it was recast as a {\em  probabilistic inference} task. In particular one defines an auxiliary probability density $p_{aux}(\bt,\bz)$, jointly on random and design variables, as follows: \footnote{For the definition of $p_{aux}$ to be valid, it suffices that $U$ is non-negative. The formulation can also account for $U$ that take negative values as long as it is  bounded from below i.e.  $U(\bt, \bz) \ge U_0 > - \infty$ ($U_0<0$), in which case one can use $U(\bt,\bz)-U_0 $ in place of $U(\bt,\bz)$}
 \be
 p_{aux}(\bt,\bz) \propto U(\bt,\bz) p_{\theta}(\bt) 
 \label{eq:aux}
\ee
The marginal $p_{aux}(\bz) \propto \int p_{aux}(\bt,\bz) ~d\bt$ is clearly proportional to $V(\bz)$. If for example one could sample  from the joint density $p_{aux}(\bt,\bz)$, the $\bs{z}-$coordinates will be  marginally distributed according to $V(\bz)$ and populate regions where this attains its maximum value(s).

The proposed reformulation allows for a uniform treatment of random $\bt$ and design variables $\bz$. More importantly, being able to infer $p_{aux}(\bz)$ (or a good approximation thereof) will not only lead to  point estimates for the maxima of the expected utility $V(\bz)$ (which coincide with the maxima of $p_{aux}(\bz)$) but also provide valuable information about the sensitivity  of the latter with respect to $\bz$ and therefore the {\em robustness} of the selected optimal design \cite{sternfels_stochastic_2011}.
Sequential Monte Carlo strategies have been previously employed \cite{amzal_bayesian-optimal_2006,kuck_smc_2006,sternfels_stochastic_2011} with significant success in identifying multiple local maxima as well as utilizing approximate, surrogate models to expedite the inference task. Nevertheless the computational cost can still be significant  as they potentially require a few thousand forward calls.

In this work we advocate an alternative probabilistic inference framework, namely  Variational Bayes (VB)
\cite{beal_variational_2003, bishop_pattern_2007}. Such methods have risen into prominence for probabilistic inference tasks in the machine learning community    \cite{jordan_introduction_1999,attias_variational_2000,wainwright_graphical_2008}.
They provide {\em approximate} inference results   by solving an optimization problem over a family of appropriately selected probability densities  with the objective of minimizing the Kullback-Leibler divergence \cite{cover_elements_1991} with  the target density (in our case $p_{aux}$). The success of such an approach  hinges upon the selection of appropriate densities that have the capacity of providing good approximations while enabling efficient (and preferably) closed-form optimization with regards to their parameters.

A pivotal role in Variational Bayesian (VB) strategies  or any other inference method, is dimensionality reduction i.e. the identification of lower-dimensional features that provide the strongest signature to the random variables and associated distributions.
Discovering a sparse set of features  has attracted great interest in many applications as in  the representation of natural images \cite{olshausen_sparse_1997} and a host of algorithms have been developed not only for finding such representations but also an appropriate dictionary for achieving this goal \cite{lewicki_learning_2000}.
   While all these tools are pertinent to the present problem they differ in a fundamental way. They are based  on several data/observations/instantiations of the vector that we seek to represent.  In our problem however we do not have such direct observations i.e. the data available pertains to the output of a model  which is nonlinearly and implicitly dependent on the vector of latent variables. Furthermore we are primarily interested in  approximating the distribution associated with  this vector rather than the dimensionality reduction itself.
   More importantly, only dimensionality reductions  that are  informative about the optimization objectives should be sought.

   A premise validated in a series of papers on the so-called ``sloppy'' models \cite{brown_statistical_2003} is that in several cases there exists a limited number of parameter combinations to which the outputs are sensitive. The overwhelming majority of directions are {\em sloppy} i.e. they embody parameter correlations that  have minor influence in the response and correspond to removable degrees of freedom. In the context of inverse problems it was found \cite{gutenkunst_universally_2007,apgar_sloppy_2010,machta_parameter_2013} that such features of the parameters can be associated with  the eigenvectors  of an appropriate Hessian or Fisher Information matrix corresponding to small eigenvalues.    
   Along these lines and by using a fully probabilistic argumentation we  develop a {\em reciprocal probabilistic PCA} \footnote{ We use the term {\em reciprocal} to distinguish from probabilistic PCA schemes \cite{tipping_probabilistic_1999} where one is interested in identifying the directions with the largest variance. In contrast, in the current setting as it will be explained later on, we are interested in the directions with lowest variance}  scheme  where eigenvectors of {\em smallest} variance are iteratively computed and are employed not only for solving the probabilistic inference problem but for identifying  the most sensitive design parameter combinations for the stochastic optimization objective.

The rest of the paper is organized as follows: The next section (Section \ref{sec:method}) presents
the essential ingredients of the  VB framework advocated, the
dimensionality reduction scheme proposed and  an iterative, coordinate-ascent algorithm that enables the identification
of all the unknowns.  Section 3 demonstrates the performance
and features of the proposed methodology in two problems from heat conduction and solid mechanics involving $\mathcal{O}(10^3)$ random and design variables.

\section{Methodology}
\label{sec:method}
As discussed in the introduction we formulate  the optimization-under-uncertainty problem as one of probabilistic inference.
To that end our goal is two-fold. Firstly, to compute  efficiently an accurate approximation of the marginal  density on the design variables $\bz$ which provides a representation of  the expected utility $V(\bz)$. Secondly, to identify a lower-dimensional subspace with regards to the design variables $\bz$ that provides an assessment of the solution's robustness by discovering  the most sensitive directions i.e. the directions along which, variations in $\bs{z}$ will cause the largest decrease in the expected utility $V(\bs{z})$. 
Such directions have been proven useful in {\em deterministic} design tasks \cite{lukaczyk_active_????}. Apart from their obvious utility, they can also facilitate the inference task  discussed previously. 
More importantly perhaps we propose a unified framework where the identification of the aforementioned lower-dimensional subspace  is performed {\em simultaneously} with the inference of the associated densities under the same Variational Bayesian objective. This yields not  only  a highly efficient algorithm (in terms of the number of forward solves) but also a highly extendable framework as discussed in the conclusions.

 We discuss first the parametrization advocated, identify latent variables and model parameters  (Section \ref{sec:param}) and subsequently demonstrate how the associated inference and learning tasks can be simultaneously addressed in the VB framework (Section \ref{sec:updates}). We finally present validation metrics that quantitatively assess the quality of the approximations derived (Section \ref{sec:val}).

\subsection{Parametrization - Dimensionality Reduction}
\label{sec:param}

Consider the auxiliary density $p_{aux}(\bt, \bz)$ defined in \refeq{eq:aux}.
This can be further extended by the introduction of an additional density $p_z(\bz)$ as follows:
\be
p_{aux}(\bt, \bz)=\frac{ U(\bt,\bz) p_{\theta}(\bt) ~p_z(\bz)}{Z}, \quad Z=\int U(\bt,\bz) p_{\theta}(\bt) ~p_z(\bz)~d\bt~d\bz
 \label{eq:aux4}
 \ee
 where $p_z(\bz)$ is the analog of the regularization term in a deterministic optimization problem. 
In many ways \refeq{eq:aux4} is a restatement of Bayes' rule with respect to $\bt,\bz$:
\be
p(\bt,\bz |data)= \frac{p(data| \bt,\bz)~p(\bt,\bz)}{p(data)}
\label{eq:bayes}
\ee
where $p_{\theta}(\bt) ~p_z(\bz)$ play the role of the {\em prior}, $U(\bt,\bz)$ is the {\em likelihood} and $p_{aux}(\bt, \bz)$ is the {\em posterior}. The connection is more apparent when one considers the utility function
 of interest in this work (\refeq{eq:util}) in which case:
\be
 p_{aux}(\bt, \bz |\bs{u}_{target} )=\frac{ e^{- \frac{1}{2}\parallel \bs{Q}^{1/2} (\bs{u}_{target}-\bs{u}(\bt, \bz))\parallel^2} p_{\theta}(\bt) ~p_z(\bz)}{Z}
 \label{eq:aux5}
 \ee
 Clearly the target response $\bs{u}_{target}$ is the direct analog of the {``data''} in \refeq{eq:bayes} and the role of marginal likelihood or model evidence term $p(data)$ is played by the normalization constant $Z$ \cite{zabaras_scalable_2008}. We make use of this connection frequently to motivate the modeling choices made, particularly with regards to the regularizations or priors which are terms that we use interchangeably.

The inference task in such a case would be formidable given the high dimensionality of $\bt, \bz$ and the cost associated with each evaluation of $U$ as previously discussed.  To address this, we propose the following decomposition of the design variables $\bz \in \RR^{d_z}$:
\be
\underbrace{\bz }_{d_z \times 1} =\underbrace{ \bs{\mu}_z}_{d_z \times 1}  + \underbrace{ \bs{W}}_{d_z  \times d_y} \underbrace{\by}_{d_y \times 1} +\underbrace{ \bez}_{d_z \times 1} 
\label{eq:red}
\ee
 The motivation behind such a decomposition is quite intuitive as it resembles a Principal Component Analysis (PCA) model \cite{tipping_probabilistic_1999}. The vector  $\bmz$ captures the central/mean value of  $\bz$, $\by$ are the reduced (and latent) coordinates 
 of $\bz$ along the linear subspace spanned by the $d_y$ columns of the matrix $\bs{W}$ and $\bez$ the residual ``noise''. As in PCA, the premise is that a few $\bs{y}$ i.e. $d_y << d_z$ suffice to capture the density of $\bz$.
 In contrast though with PCA where the reduced coordinates are associated with the principal directions of largest variance, the $\by$ employed here should do the exact opposite i.e. identify directions with smallest variance that imply largest sensitivity.  We explain this in more detail in the next Section.

 The linear decomposition of a high-dimensional vector such as $\bz$ has received a lot of attention in several different fields. Most commonly $\bz$ represents a high-dimensional signal  (e.g. an image, an audio/video recording)  and $\bs{W}$ consists of  an over- or under-complete basis set   \cite{olshausen_sparse_1997,dobigeon_bayesian_2010} 
 which attempts to encode the signal as {\em sparsely} as possible. Significant advances in Compressed Sensing \cite{candes_robust_2006} or Sparse Bayesian Learning \cite{wipf_sparse_2004} have been achieved in recent years along these lines. A host of deterministic \cite{lee_efficient_2006} or probabilistic \cite{seeger_variational_2010} algorithms have been developed for identifying the reduced-coordinates $\by$  as well as techniques for learning the most appropriate set of basis $\bs{W}$  (dictionary learning) i.e. the one that can lead to the sparsest possible 
representation.  

We adopt a simpler representation for the input random variables $\bt \in \RR^{\dt}$:
\be
\underbrace{\bt }_{\dt \times 1} =\underbrace{ \bmt }_{\dt \times 1}  +\underbrace{ \bet}_{\dt \times 1} 
\label{eq:redtheta}
\ee
the usefulness of which will become apparent in the sequel.
 In a fully probabilistic setting all the aforementioned parameters ($\by,\bez, \bet, \bmz,\bs{W},\bmt$) and the corresponding densities arising from \refeq{eq:aux4} would be sought. Such an inference problem would in general be formidable particularly with regards to $\bmz,\bs{W},$ whose dimension is dominated by $d_z >>1$. To {address} this difficulty we propose computing  point estimates for $\bmz,\bs{W},\bmt$ while quantifying the appropriate densities for  $\by,\bez, \bet$. 
 We distinguish therefore  between:
\bi
\item the {\em latent variables} $\by,\bez, \bet$.
\item and {\em model parameters} $\bs{R}=\{ \bmz,\bs{W},\bmt\}$.
\ei
The computation of appropriate distributions for the latent variables and point estimates for $\bs{R}$  will be addressed simultaneously under the VB framework discussed in the sequel.

\subsection{ Variational Bayesian approximation}

Given the re-parametrization of the primal variables $\bt,\bz$ in Equations (\ref{eq:red}), (\ref{eq:redtheta}), one can write the target auxiliary density as:
\be
 p_{aux}(\by,\bez,\bet, \bs{R})= \frac{U(\bmt+\bet, \bmz+\bs{W}\by+\bez) p_{\theta}(\bmt+\bet) p_y(\by) p_{\eta_z}(\bez) p_{\mu_z}(\bmz) p_{W}(\bs{W}) }{Z}
 \label{eq:auxjoint}
\ee
where in place of the regularization $p_z(\bz)$ on $\bz$ we employ regularizations (priors) on the corresponding parameters $\by,\bez$ and $\bmz,\bs{W}$ (\refeq{eq:red}).
As discussed earlier rather than approximating the whole $p_{aux}$ which would pose significant difficulties, we seek point estimates for $\bs{R}$ by maximizing the (marginal) density $p_{aux}(\bs{R} )$:
\be
p_{aux}(\bs{R})= \int p_{aux}(\by,\bez,\bet, \bs{R}) ~d\by ~d\bez ~d\bet
\label{eq:auxR}
\ee
Such a maximization would amount to an analog of Maximum-A-Posteriori (MAP) estimates in a Bayesian setting.

To that end, for any density $q(\by,\bez, \bet)$  on the latent variables and  by employing Jensen's inequality we obtain that \cite{wainwright_graphical_2008}:
\be
\begin{array}{ll}
 \log p_{aux}(\bs{R}) & = \log \int p_{aux}(\by,\bez,\bet, \bs{R}) ~d\by ~d\bez ~d\bet \\
 & = \log \int q(\by,\bez, \bet) \frac{ p_{aux}(\by,\bez,\bet, \bs{R})}{q(\by,\bez, \bet)}  ~d\by ~d\bez ~d\bet \\
 & \ge \int q(\by,\bez, \bet) \log  \frac{ p_{aux}(\by,\bez,\bet, \bs{R})}{q(\by,\bez, \bet)}  ~d\by ~d\bez ~d\bet \\
 & = \mathcal{F}(q(\by,\bez, \bet), \bs{R})
\end{array}
\label{eq:varF}
\ee

The variational lower bound $\mathcal{F}$ given above has an intimate connection with the  KL-divergence between $q(\by,\bez, \bet)$ and the {\em conditional } density $p_{aux}(\by,\bez,\bet | \bs{R}) = \frac{p_{aux}(\by,\bez,\bet, \bs{R})}{ p_{aux}(\bs{R}) }$ which can be expressed as:
\be
\begin{array}{ll}
 0 \le KL(q(\by,\bez, \bet)|| p_{aux}(\by,\bez,\bet | \bs{R})) & = -E_q\left[ \log \frac{ p_{aux}(\by,\bez,\bet | \bs{R})}{q(\by,\bez, \bet)} \right] \\
  & = - E_q\left[\frac{ p_{aux}(\by,\bez,\bet, \bs{R})}{p_{aux}(\bs{R}) q(\by,\bez, \bet)}  \right] \\
  & = \log p_{aux}(\bs{R}) -\mathcal{F}(q(\by,\bez, \bet), \bs{R})
\end{array}
\label{eq:kl}
\ee
We note that when $q(\by,\bez, \bet) \equiv  p_{aux}(\by,\bez,\bet | \bs{R})$ the KL-divergence attains its minimum value $0$, while  $\mathcal{F}$ attains its maximum value  with respect to $q$ (given $\bs{R}=(\bmz,\bs{W},\bmt)$) and becomes equal to $\log p_{aux}(\bs{R})$. 
On the other hand the poorer  the approximation that $q(\by,\bez, \bet)$ provides to $p_{aux}(\by,\bez,\bet | \bs{R})$, the larger the KL-divergence and the smaller $\mathcal{F}$ (as a function of $q$) becomes.

\begin{figure}[!t]{
        \centering
        \psfrag{ftm1}{\small $\mathcal{F}(q^{(t-1)},\bs{R}^{(t-1)})$}
        \psfrag{kltm1}{\small $KL(q^{(t-1)} || p_{aux}(. |\bs{R}^{(t-1)}))$}
         \psfrag{logptm1}{\small $\log p_{aux}(\bs{R}^{(t-1)})$}
         \psfrag{estep}{VB-E-step}
  \psfrag{ft}{\small $\mathcal{F}(q^{(t)},\bs{R}^{(t-1)})$}
        \psfrag{klt}{\small $KL(q^{(t)} || p_{aux}(. |\bs{R}^{(t-1)}))$}
         \psfrag{logpt}{\small$\log p_{aux}(\bs{R}^{(t-1)})$}
         \psfrag{mstep}{VB-M-step}
  \psfrag{ftp1}{\small $\mathcal{F}(q^{(t)},\bs{R}^{(t)})$}
        \psfrag{kltp1}{\small $KL(q^{(t)} || p_{aux}(. |\bs{R}^{(t)}))$}
         \psfrag{logpt1}{\small $\log p_{aux}(\bs{R}^{(t)})$}
      \includegraphics[width=0.90\textwidth]{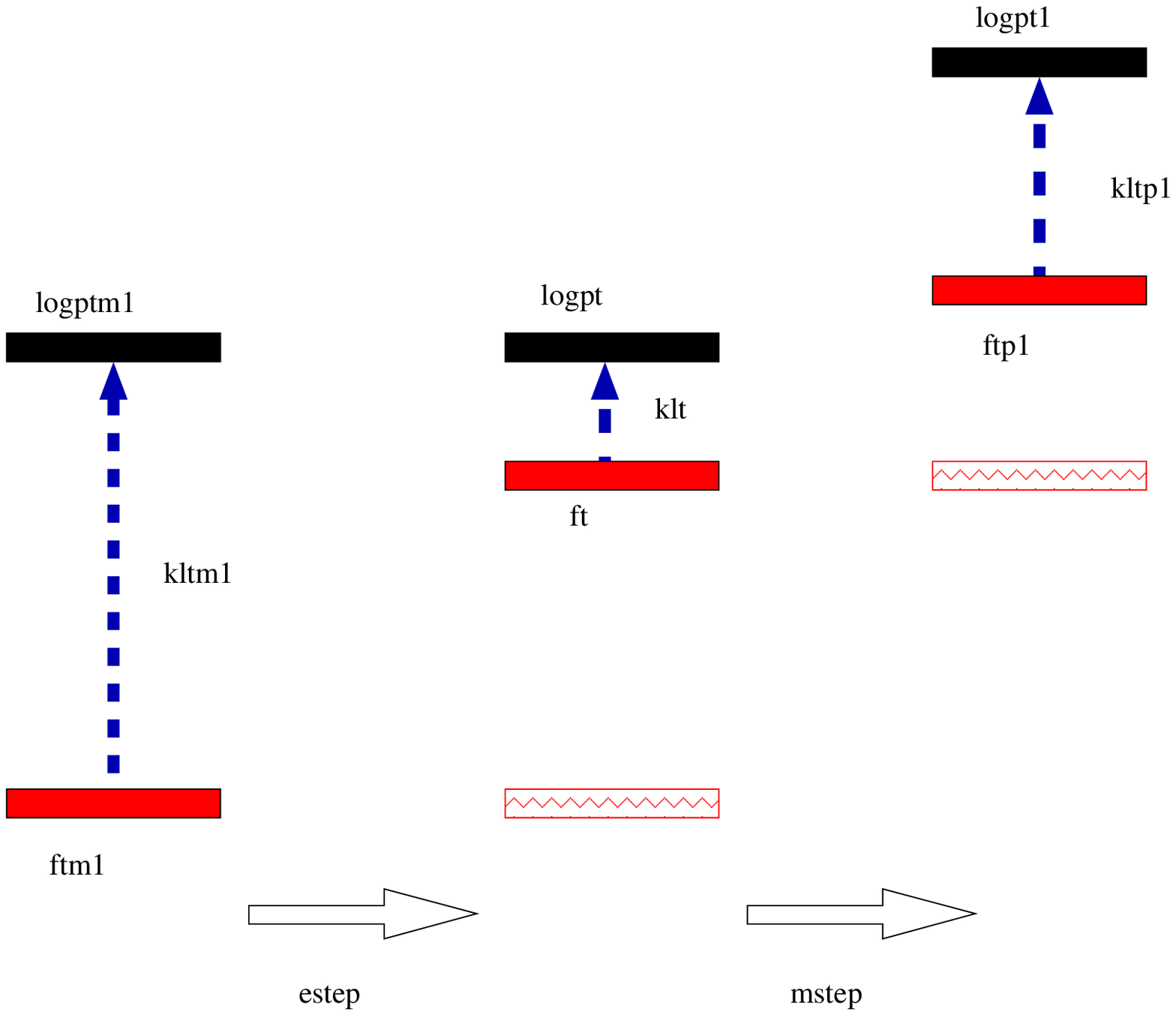}}
      \caption{ During the VB-E step, optimization with respect to the approximating distribution $q$ takes place, whereas during the VB-M step, $\mathcal{F}$ is optimized with respect to the model parameters $\bs{R}$ (adapted from \cite{beal_variational_2003-2})}
       \label{fig:vbem}
\end{figure}

The aforementioned discussion suggests an iterative optimization scheme that resembles the  Variational Bayes - Expectation-Maximization (VB-EM) methods that have  appeared  in Machine Learning literature  \cite{beal_variational_2003}. 
At each iteration $t$, one alternates between (Figure \ref{fig:vbem}):
\bi
\item \textbf{VB-Expectation}: Given $\bs{R}^{(t-1)}$, find:
\be
q^{(t)} (\by,\bez, \bet) =\arg \max_q \mathcal{F}(q(\by,\bez, \bet), \bs{R}^{(t-1)})
\label{eq:vbe}
\ee
\item \textbf{VB-Maximization}: Given $q^{(t)} (\by,\bez, \bet)$, find:
\be
\bs{R}^{(t)} =\arg \max_R \mathcal{F}(q^{(t)}(\by,\bez, \bet), \bs{R})
\label{eq:vbm}
\ee
\ei

\begin{figure}[!t]{
        \centering
        \psfrag{q}{$q(\by,\bez, \bet)$}
        \psfrag{R}{$\bs{R}=\{\bmz,\bs{W},\bmt\}$}
                \psfrag{F}{$\mathcal{F}(q(\by,\bez, \bet),\bs{R})$}
\includegraphics[width=0.60\textwidth]{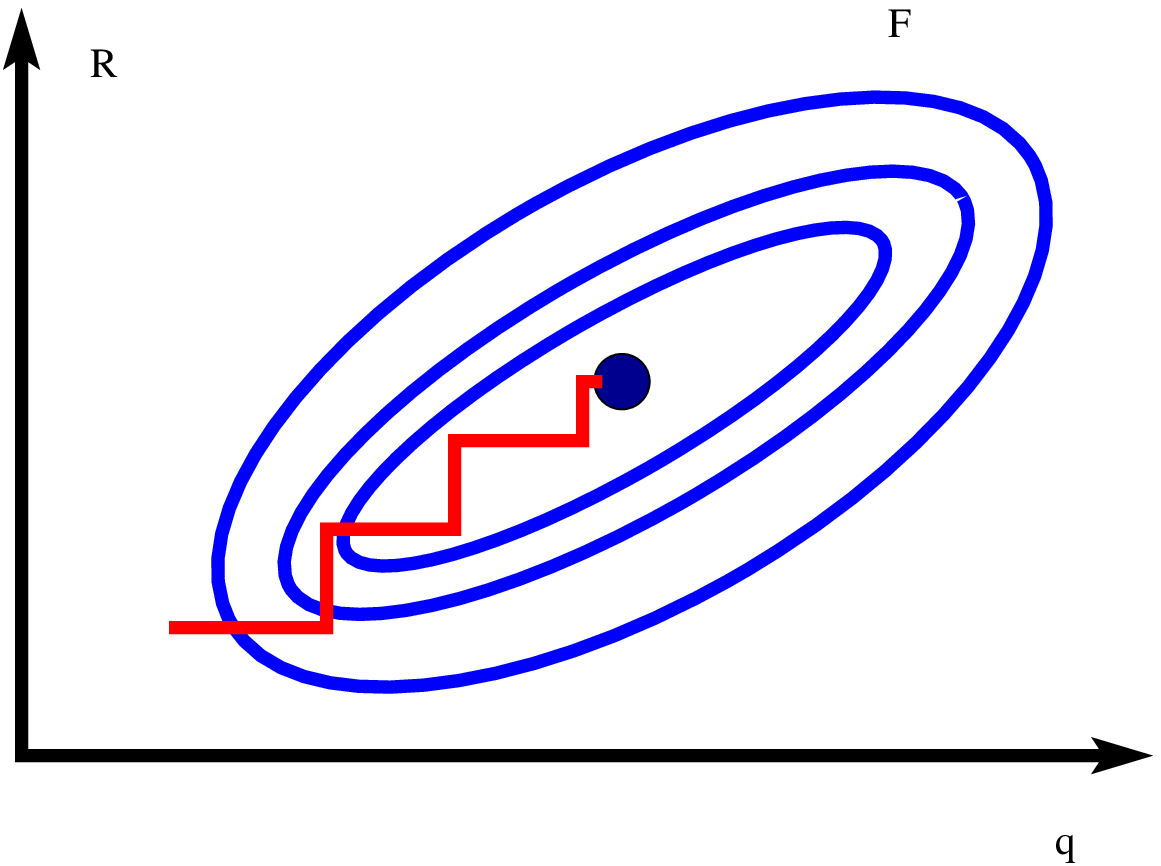}}
        \caption{Variational Bayesian Expectation-Maximization (VB-EM, \cite{beal_variational_2003})}
        \label{fig:maxF}
\end{figure}

In plain terms, the strategy advocated in order to carry out the inference task explained can be described as a generalized coordinate ascent with regards to $\mathcal{F}$ (Figure \ref{fig:maxF}).

%
%

%
%

\subsection{Approximations}

The variational lower bound $\mathcal{F}$ (\refeq{eq:varF}) is the objective function in the proposed scheme.
In this Section we discuss its form for the utility function of interest (\refeq{eq:util}) and an  isotropic $\bs{Q}=\tau_Q \bs{I}_n$. Furthermore we discuss necessary approximations that enable the VB-EM steps.
We defer discussions on the validity and quantitative assessment of these approximations for Section \ref{sec:val}.

In particular, from Equations (\ref{eq:auxjoint}) and (\ref{eq:varF}), we have:
\be
\begin{array}{ll}
\mathcal{F}(q(\by,\bez, \bet), \bs{R}) & = \int q(\by,\bez, \bet) \log  \frac{ p_{aux}(\by,\bez,\bet, \bs{R})}{q(\by,\bez, \bet)}  ~d\by ~d\bez ~d\bet \\
& = E_q\left[ \frac{ U(\bmt+\bet, \bmz+\bs{W}\by+\bez) p_{\theta}(\bmt+\bet) p_y(\by) p_{\eta_z}(\bez) p_{\mu_z}(\bmz) p_{W}(\bs{W}) }{Z~q(\by,\bez, \bet)} \right] \\
& =  E_q\left[ U(\bmt+\bet, \bmz+\bs{W}\by+\bez)\right]+E_q\left[\frac{ p_{\theta}(\bmt+\bet) p_y(\by) p_{\eta_z}(\bez)}{ q(\by,\bez, \bet)} \right] \\ 
 & +\log p_{\mu_z}(\bmz)+\log p_{W}(\bs{W}) \\
 & = \mathcal{F}_U+\mathcal{F}_{reg} +\log p_{\mu_z}(\bmz)+\log p_{W}(\bs{W}) \\
\end{array}
\label{eq:ftot}
\ee
where we distinguish the individual terms:
\be
\mathcal{F}_U= E_q\left[ U(\bmt+\bet, \bmz+\bs{W}\by+\bez)\right]
\label{eq:fu}
\ee
\be
\mathcal{F}_{reg}=E_q\left[\frac{ p_{\theta}(\bmt+\bet) p_y(\by) p_{\eta_z}(\bez)}{ q(\by,\bez, \bet)} \right]
\label{eq:freg}
\ee
We note that in the aforementioned expressions we omit $\log Z$ as this does not depend on $q$ nor $\bs{R}$ and therefore does not affect any of the VB-EM results.

Furthermore, we note that:
\be
\begin{array}{ll}
 \mathcal{F}_U & = E_q\left[ U(\bmt+\bet, \bmz+\bs{W}\by+\bez)\right] \\
 & = -\frac{\tau_Q}{2} E_q[|\bs{u}_{target}-\bs{u}(\bmt+\bet, \bmz+\bs{W}\by+\bez)|^2]
\end{array}
\ee
is not only analytically intractable but also poses  significant difficulties due to the computational expense associated with  each forward call for the evaluation of  $\bs{u}(\bmt+\bet, \bmz+\bs{W}\by+\bez)$. 
To alleviate these issues we propose a  {\em linearization} of the output vector $\bs{u}$ around for $\bt=\bmt$ and $\bz=\bmz$. In particular:
\be
\begin{array}{ll} 
\bs{u}(\bmt+\bet, \bmz+\bs{W}\by+ \bez) & \approx \bs{u}(\bmt, \bmz)+\bgt \bet+\bgz (\bs{W}\by+ \bez) \\
& = \bs{u}(\bmt, \bmz)+\bgt \bet+\bgz \bs{W}\by+ \bgz \bez
\end{array}
\label{eq:linu}
\ee
where $\bgt=\frac{\pa \bs{u}}{\pa \bt}, \bgz=\frac{\pa \bs{u}}{\pa \bz}$ evaluated at $(\bmt,\bmz)$. These derivatives can be computed using adjoint formulations when the forward model is a system of PDEs as in the examples considered in Section 3.
Such a linearization will lead to a quadratic, Gauss-Newton-type,  expression upon  substitution in the log-utility function:
\be
\begin{array}{ll}
 |\bs{u}_{target}-\bs{u}(\bmt+\bet, \bmz+\bs{W}\by+ \bez)|^2 & \approx |\bs{u}_{target}-\bs{u}(\bmt, \bmz)|^2+\bet^T \bgt^T \bgt \bet \\
 & + \by^T \bs{W}^T \bgz^T \bgz \bs{W} \by+\bez^T \bgz^T \bgz \bez \\
 & - 2(\bs{u}_{target}-\bs{u}(\bmt, \bmz))^T ( \bgt \bet+\bgz \bs{W}\by+ \bgz \bez) \\
 & + 2\bet^T \bgt^T \bgz \bs{W} \by+ \bez^T \bgz^T ( \bgt \bet+\bgz \bs{W}\by)
\end{array}
\label{eq:linu2}
\ee
We note here that a quadratic approximation could also be obtained using a $2^{nd}$ order Taylor series expansion of $|\bs{u}_{target}-\bs{u}(\bmt+\bet, \bmz+\bs{W}\by+ \bez)|^2$ directly. This would require the computation of the Hessian matrix which is also possible using adjoint formulations albeit at a significant additional cost \cite{biegler_large-scale_2003}. Furthermore, for very large $d_{\theta}, d_z>>1$ the storage of the Hessian might be impractical. The reason for the quadratic approximation advocated is that it leads to closed-form expressions for the density $q$ in the VB-Expectation step (\refeq{eq:vbe}) as it will become apparent in Section \ref{sec:updates}. Higher-order approximations would also be suitable as long as the latter requirement is satisfied.

A  quadratic expression can be obtained by a $2^{nd}$-order Taylor series expansion of   $\log p_{\theta}(\bmt+\bet)$ around $\bmt$ without significant cost (most often than not, analytically) i.e.:
\be
\log p_{\theta}(\bmt+\bet) \approx \log p_{\theta}(\bmt)+\bet^T \frac{\pa \log p_{\theta}}{\pa \bt}|_{\bmt} +\frac{1}{2} \bet^T \frac{\pa \log p_{\theta}}{\pa \bt\pa  \bt^T}|_{\bmt} \bet
\ee
In the case that $p_{\theta}(\bt)$ is a multivariate Gaussian as in the examples considered i.e. $\mathcal{N}(\bs{\mu}_{\theta 0}, \bs{C}_{\theta 0})$, then the quadratic expression is exact and attains the form:
\be
\begin{array}{ll}
 \log p_{\theta}(\bmt+\bet)=-\frac{1}{2}(\bmt+\bet-\bs{\mu}_{\theta 0})^T \bs{C}_{\theta 0}^{-1}(\bmt+\bet-\bs{\mu}_{\theta 0})
 \label{eq:linp}
\end{array}
\ee

\subsection{Prior specification for latent variables and model parameters}
\label{sec:priors}

 The latent, reduced coordinates $\by \in \RR^{d_y}$ capture the variation of $\bz$ around its mean $\bmz$ along the directions of $\bs{W}$ as implied by \refeq{eq:red}. It is therefore reasonable to assume that, a priori, these should have zero mean and should be uncorrelated \cite{tipping_probabilistic_1999}. For that purpose we adopt a multivariate Gaussian prior (denoted by $p_y(\by)$ in the Equations of the previous section) with a diagonal covariance denoted by $\bs{C}_{y0}=diag(\tau_{0,i}^{-1}), i=1, \ldots d_y$. In the examples presented in Section \ref{sec:numericalexamples},  $\tau_{0,i}$ are  set to the same value $\tau_{y0}$ i.e.:
 \be
 \bs{C}_{y0}=\tau_{y0}^{-1} \bs{I}_{d_y}
 \label{eq:priory}
 \ee
 Alternatively, one can select $\tau_{0,1}^{-1} <   \tau_{0,2}^{-1}< \ldots \tau_{0,d_y}^{-1}$ which induces a stochastic ordering of the reduced coordinates $\by$  since $\bz$ is invariant to permutations of the entries of the $\bs{y}$ and the columns of $\bs{W}$ (\refeq{eq:red}). %
 
 The remaining latent variables $\bez$ account for the part of $\bz$ that is not captured by $\bmz+\bs{W}\by$ (\refeq{eq:red}) and should therefore account for  the variance in the subspace orthogonal to $\bs{W}$. 
 {\em The premise in the formulation advocated (Figure \ref{fig:sloppy}) is that $\by$ capture the most sensitive directions (locally) around $\bmz$ which are much smaller in number than the dimensionality of $\bz$ i.e. $d_y<< d_z$. The variance of $\by$ should therefore be \textbf{the smallest} amongst all possible directions.
 The remaining directions where the variance is much larger should be captured by $\bez$. We use therefore an isotropic Gaussian as a ``prior`` for $\bez$ i.e. $p_{\eta_z}(\bez)$ is  $\mathcal{N}(\bs{0}, \tau_{z0}^{-1} (\bs{I}-\bs{W}\bs{W}^T))$ \footnote{The covariance $(\bs{I}-\bs{W}\bs{W}^T)$ is obviously improper as it has $d_y$ zero eigenvalues to reflect the fact that $\bez$ is inherently $(d_z-d_y)$-dimensional}
 where the prior variance is set much larger than the prior variances of $\by$ e.g.:
 \be
 \tau_{z0}^{-1}= \frac{\tau_{y0}^{-1}}{\epsilon^2}
 \label{eq:priorez}
 \ee
 where $\epsilon^2 <<1$ (Section \ref{sec:numericalexamples}). 
 We point out that this is the premise invoked also in the context of Sloppy Models \cite{gutenkunst_universally_2007,apgar_sloppy_2010,machta_parameter_2013},  whose behavior depends only on a few stiff combinations of parameters (accounted here by $\bs{W}$ and $\by$), with many sloppy parameter directions largely unimportant for model predictions (accounted here by $\bez$).
  We  also note here the fundamental difference with PCA decompositions which attain the same form as $\refeq{eq:red}$. In PCA, $\bs{W}$ and the latent variables $\bs{y}$ capture the directions of \textbf{largest} variance and $\bez$ account for the remaining variance  which is isotropic, smaller, and  superimposed on the directions $\bs{W}$. 
 }
 

 With regards to the regularization (prior) specification $p_{W}(\bs{W})$ on $\bs{W}$ we note that its  $d_y$ columns $\bs{w}_i, ~i=1,\ldots 
 d_y$ span the subspace over which an approximation of $\bz$ is sought. 
 We note  that $\bz$ depends on the product $\bs{W} \by$ which would remain invariant by appropriate rescaling of each pair of $\bs{w}'_i=\alpha_i ~\bs{w}_i$ and $y'_i=\frac{1}{\alpha_i} y_i$ for any $\alpha_i$. Hence, to resolve identifiability issues we require that $\bs{W}$ is {\em orthogonal} i.e. $\bs{W}^T \bs{W}=\bs{I}_{d_y}$ where $\bs{I}_{d_y}$ is the $d_y-$dimensional identity matrix. This is equivalent to employing a uniform prior on $\bs{W}$ on the Stiefel manifold $V_{d_y}(\RR^{d_z})$ \cite{muirhead_aspects_1982}. 
 
 The final aspect of the prior model pertains to $p_{\mu_z}(\bs{\bmz})$. As this is closely related to the physical meaning of the design variables $\bz$ we make this specific for each of  the examples considered in Section 3.

 \subsection{VB- Expectation-Maximization:  Update equations for $q(\bet,\by,\bez)$ and $\bs{R}$}
 \label{sec:updates}
 
 We consider Gaussian families of approximating distributions $q(\bet,\by,\bez)$ with the following mean and covariance characteristics:
 \be
 \left[
 \begin{array}{l}
  E_q[\bet]=\bs{0} \\ E_q[\by]=\bs{0} \\ E_q[\bez]=\bs{0}
 \end{array}
 \right],
 \quad
 \left[
 \begin{array}{ccc}
 E_q[\bet \bet^T]= \bs{C}_{\theta \theta} &  E_q[\bet \by^T]=\bs{C}_{\theta y} & E_q[\bet \bez^T]= \bs{0} \\
 E_q[\by \bet ^T]=\bs{C}_{\theta y}^T & E_q[\by \by^T]=\bs{C}_{yy} &  E_q[ \by\bez^T]=\bs{0} \\
 E_q[\bez \bet^T]= \bs{0} & E_q[\bez  \by^T]=\bs{0} &  E_q[ \bez \bez^T]=\tau_z^{-1} ( \bs{I} -\bs{W}\bs{W}^T) 
 \end{array} \right] 
 \label{eq:postq}
 \ee
  This form is postulated for the following reasons:
  \bi
  \item $\bet$ expresses variations of $\bt$ from its mean $\bmt$ (\refeq{eq:redtheta}) and should   therefore have a mean zero.
 \item $\by$ and $\bez$ express variations of $\bz$ from its mean $\bmz$ (\refeq{eq:red})  and should also have a mean zero. 
  \item $\bez$ expresses residual variation (noise) of $\bz$ from $\bmz+\bs{W}\by$. Apart from having a zero mean, it is assumed to be uncorrelated with  $\bet$  as well as $\by$.
  \item $\bez$ accounts for variance in the subspace orthogonal to $\bs{W}$. Along this it is assumed that the variance is isotropic and equal to $\tau_z^{-1}$ (to be determined)
    \ei
  
  Before embarking in the presentation of the expressions for the aforementioned parameters, we note that $q$ provides an approximation to $p_{aux}$ and therefore its marginal with respect to $\by, \bez$ can be used to approximate the marginal on $\bz$ i.e. $p_{aux}(\bz)$ which is proportional to the expected utility $V(\bz)$. We note  that based on \refeq{eq:postq}, the marginal $q(\by,\bez)$ will also be a Gaussian and there the approximate $p_{aux}(\bz)$ will be a Gaussian with the following mean and covariance:
  \be
  E[\bz]=\bmz, \quad E[\bz \bz^T]=\bs{C}_{zz}=\bs{W} \bs{C}_{yy} \bs{W}^T+\tau_{z}^{-1} ( \bs{I} -\bs{W}\bs{W}^T) 
  \label{eq:postz}
  \ee
  We can therefore approximate (up to a multiplicative constant) the expected utility $V(\bz)$ as:
  \be
  V(\bz) \approx V(\bmz+\bs{W}\by+\bez) \propto e^{-\frac{1}{2} (\bz-\bmz)^T \bs{C}_{zz}^{-1} (\bz-\bmz)}
  \label{eq:Vapprox}
  \ee
  
\begin{figure}[!t]
 \centering
  \psfrag{s1}{$\sigma_1^2$}
    \psfrag{s2}{$\sigma_2^2$}
  \psfrag{z1}{$z_1$}
  \psfrag{z2}{$z_2$}
  \psfrag{mz}{$\bmz$}
  \psfrag{W}{$\bs{W}$}
  \psfrag{Wo}{$\bs{W}^{\perp}$}
\psfrag{stiff}{\rotatebox{35}{sensitive/stiff}}
\psfrag{sloppy}{\rotatebox{35}{insensitive/sloppy}}
  \includegraphics[width=.85\textwidth]{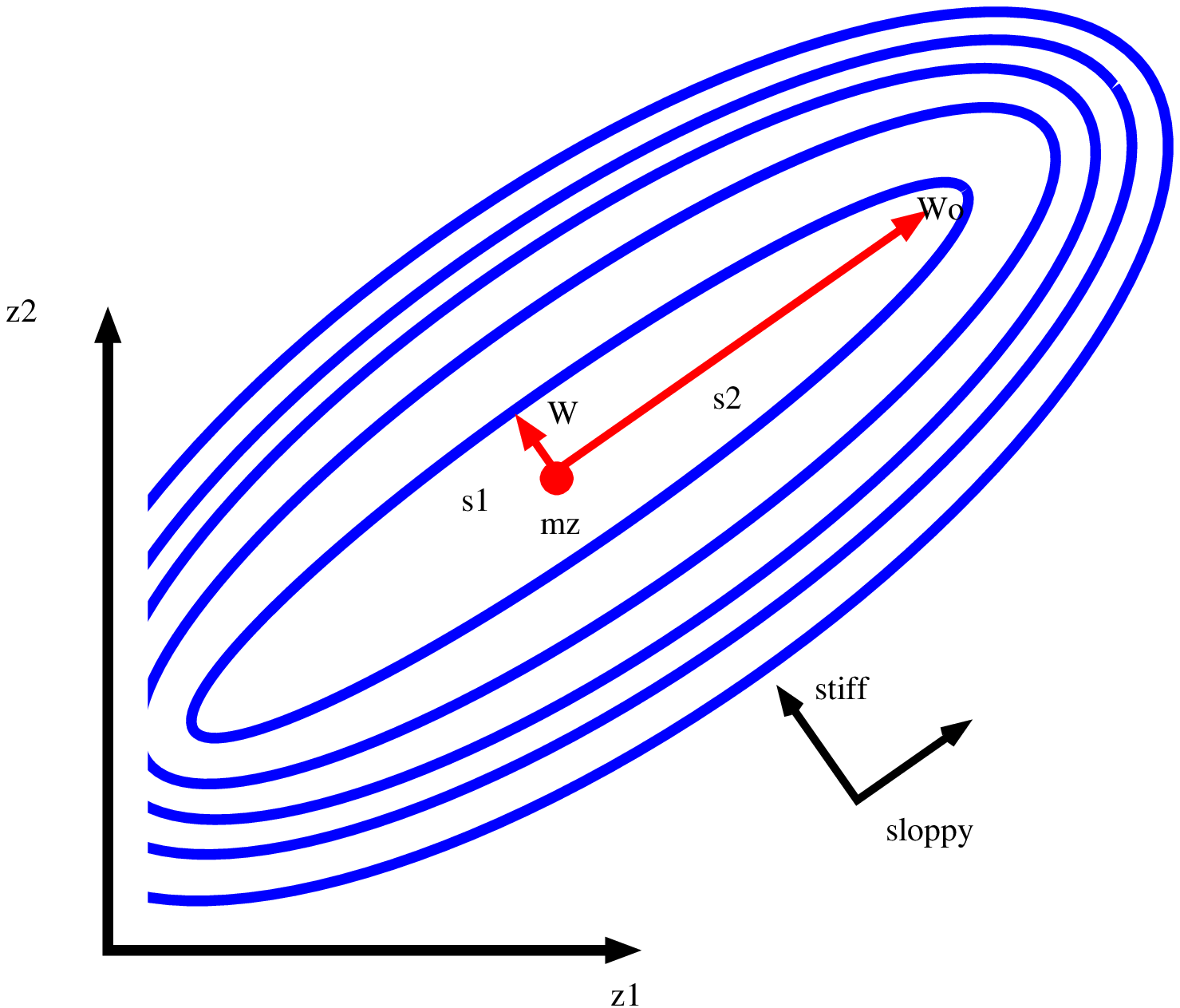}
  \caption{Illustration in two dimensions ($d_z=2$) of $p_{aux}(\bz)$ (which approximates expected utility $V(\bz)$) in the vicinity of (local) maximum $\bmz$. The most sensitive/stiff directions are captured by $\bs{W}$ and their variance $\sigma_1^2$ is accounted by $\bs{y}$. The most insensitive/sloppy directions are orthogonal to $\bs{W}$ (along $\bs{W}^{\perp}$) and their variance $\sigma_2^2 >> \sigma_1^2$ is accounted by $\bez$.  }
  \label{fig:sloppy}
 \end{figure}

  Let $\hat{\bs{w}}_j, ~j=1,\ldots, d_z$ denote the eigenvectors of $\bs{C}_{zz}$ and $\sigma_1 < \sigma_2^2 < \ldots < \sigma_{d_z}^2$ the corresponding eigenvalues in ascending order (Figure \ref{fig:sloppy}).
  Consider (local) variations $\Delta \bz_j$ of $\bz$ from $\bmz$ along the distinct directions  $\hat{\bs{w}}_j$ i.e.:
  \be
  \Delta \bz_j=\bz-\bmz=~\alpha~\hat{\bs{w}}_j 
  \ee
  Then:
  \be
  V(\Delta \bz_1)\propto e^{-\frac{1}{2} \frac{\alpha^2}{\sigma_1^2} } < V(\Delta \bz_2)\propto e^{-\frac{1}{2} \frac{\alpha^2}{\sigma_2^2} } < \ldots <V( \Delta \bz_{d_z})\propto e^{-\frac{1}{2} \frac{\alpha^2}{\sigma_{d_z}^2} } 
  \ee
  Hence the expected utility of  competing designs $z_j$ will decrease faster for variations along directions with the {\em smaller} variances/eigenvalues which represent the directions of higher sensitivity.
  
 The methodology developed is based on the postulate that the number of {\em sensitive directions} is small compared to $d_z$ and can be captured by the $d_y <<d_z$ latent variables $\by$ and the vectors in $\bs{W}$. Most of the remaining directions are assumed to be sloppy i.e. have a much higher variance and can be represented by $\bez$.
 The $d_y$ sensitive directions can be found by diagonalizing $\bs{C}_{yy}=\bs{U}~diag(\sigma_{1\div d_y}^2)~\bs{U}^T$ and as a result:
 \be
 \bs{\hat{W}}=\left[ \begin{array}{cccc} \bs{\hat{w}}_1 & \bs{\hat{w}}_2 & \ldots & \bs{\hat{w}}_{d_y} \end{array} \right] =\bs{W} \bs{U}
 \label{eq:what}
 \ee
 
%
%
%
  
  From the expressions of $\mathcal{F}_U, \mathcal{F}_{reg}$ in Equations (\ref{eq:ftot}) (\ref{eq:fu}), (\ref{eq:freg}) and the approximations in Equations (\ref{eq:linu2}) and (\ref{eq:linp}), we obtain (up to additive constants):
  \be
  \begin{array}{ll}
\mathcal{F}(q(\by,\bez, \bet), \bs{R}) = &-\frac{\tau_Q}{2} \left( |\bs{u}_{target}-\bs{u}(\bmt, \bmz)|^2 \right. \hfill (\textrm{from $E_q[U]$}) \\
&~ +\bgt^T \bgt: \bs{C}_{\theta \theta} +\bs{W}^T \bgz^T \bgz \bs{W}: \bs{C}_{yy}+\tau_z^{-1} \bgz^T \bgz:(\bs{I}-\bs{W}\bs{W}^T) \\
& ~\left. +2  \bgt^T \bgz \bs{W} : \bs{C}_{\theta y} \right) \\
& -\frac{1}{2}(\bmt-\bs{\mu}_{\theta 0})^T \bs{C}_{\theta 0}^{-1}(\bmt-\bs{\mu}_{\theta 0})  \hfill (\textrm{from $E_q[p_{\theta}]$})\\
& -\frac{1}{2}  \bs{C}_{\theta 0}^{-1}:  \bs{C}_{\theta \theta} \\
& -\frac{\tau_{y0}}{2} \bs{I}:\bs{C}_{yy} \hfill (\textrm{from $E_q[p(\by)]$}) \\
& -\frac{ d_z-d_y}{2} \frac{\tau_{z0}}{\tau_{z}} \hfill (\textrm{from $E_q[p(\bez)]$}) \\
& +\frac{1}{2} \log  \left| \begin{array}{cc} \bs{C}_{\theta \theta} & \bs{C}_{\theta y} \\ \bs{C}_{\theta y}^T & \bs{C}_{yy} \end{array} \right| \hfill (\textrm{from $E_q[q(\bet,\by)]$})\\
& + \frac{d_z-d_y}{2} \log \tau_z \hfill (\textrm{from $E_q[q(\bez)]$}) \\
&  +\log p_{\mu_z}(\bmz) +\log p_{W}(\bs{W})
  \end{array}
  \label{eq:Fopt}
\ee

The iterative VB-EM optimization with regards to $q$ and $\bs{R}$ (Figure \ref{fig:maxF}) proceeds then as follows:
\bi
\item \textbf{VB-Expectation}: Given the current $\bs{R}$ find the optimal $q$ (i.e. the optimal $\bs{C}_{\theta \theta}, \bs{C}_{\theta y}, \bs{C}_{yy}, \tau_z$):
\be
\left[ \begin{array}{cc} \bs{C}_{\theta \theta}^{opt} & \bs{C}_{\theta y}^{opt} \\ sym. & \bs{C}_{yy}^{opt} \end{array} \right]^{-1} =
\left[ \begin{array}{cc}  \tau_Q \bgt^T \bgt+\bs{C}_{\theta 0}^{-1} &  \tau_Q \bgt^T \bgz \bs{W} \\
    sym. & \tau_Q \bs{W}^T\bgz^T \bgz \bs{W}+\tau_{y0} \bs{I}
\end{array} \right]
\label{eq:optq}
\ee
and:
\be
\tau_z^{opt}=\tau_{z0}+\frac{1}{d_z-d_y} \tau_Q \bgz^T \bgz : (\bs{I}-\bs{W}\bs{W}^T)
\label{eq:opttau}
\ee

\item \textbf{VB-Maximization}: Given the current $q$ (i.e.  $\bs{C}_{\theta \theta}, \bs{C}_{\theta y}, \bs{C}_{yy}, \tau_z$), find the optimal $\bs{R}=\{ \bmz,\bs{W},\bmt\}$. To carry out this task, it suffices to consider only the terms of $\mathcal{F}$  in \refeq{eq:Fopt}, that depend on the parameters of interest i.e.:
\be
(\bmz^{opt},\bmt^{opt})=\arg \max_{\bmz,\bmt} \mathcal{\hat{F}}_{\mu}(\bmz,\bmt), \quad \bs{W}^{opt}=\arg \max_{\bs{W}} \mathcal{\hat{F}}_{W}(\bs{W})
\ee
where:
\be
\begin{array}{ll}
\mathcal{F}_{\mu}(\bmz,\bmt) = &-\frac{\tau_Q}{2} \left( |\bs{u}_{target}-\bs{u}(\bmt, \bmz)|^2 \right)  \\
& -\frac{1}{2}(\bmt-\bs{\mu}_{\theta 0})^T \bs{C}_{\theta 0}^{-1}(\bmt-\bs{\mu}_{\theta 0}) +\log p_{\mu_z}(\bmz)
\end{array}
\label{eq:optmu}
\ee
and:
\be
\begin{array}{ll}
\mathcal{F}_{W}(\bs{W})& = -\frac{\tau_Q}{2} \bs{W}^T \bgz^T \bgz \bs{W}: \bs{C}_{yy}  -\frac{\tau_Q}{2} \tau_z^{-1} \bgz^T \bgz:(\bs{I}-\bs{W}\bs{W}^T)  \\
 &- \frac{\tau_Q}{2}  2\bgt^T \bgz \bs{W} : \bs{C}_{\theta y}   +\log p_{W}(\bs{W}) \\
 & = -\frac{\tau_Q}{2} \bs{W}^T \bgz^T \bgz \bs{W}: (\bs{C}_{yy}-\tau_z^{-1} \bs{I}) \\
 & - \frac{\tau_Q}{2}  2\bgt^T \bgz \bs{W} : \bs{C}_{\theta y}+\log p_{W}(\bs{W})
\end{array}
\label{eq:objW}
\ee
\ei

Some remarks are warranted at this stage:
\bi
\item The maximization  of $\mathcal{F}_{\mu}$ with respect to $(\bmt,\bmz)$  can be carried out using any nonlinear optimization scheme. We discuss in   \ref{app:mu} a Gauss-Newton-type scheme which requires only first-order derivatives of $\bs{u}$.  We note that this is the {only part} of the VB-EM scheme proposed that requires calls to the forward solver for the computation of $\bs{u}$ and its derivatives. Hence in the context of large-scale, complex models this step controls, to a large extent, the overall cost of the proposed algorithm.

\item The updates of $(\bmt,\bmz)$ in \refeq{eq:optmu} are  {\em decoupled} from the rest i.e. $q$ and $\bs{W}$. This is a direct consequence of the assumption on $q$ that $E_q[\by]=E_q[\bez]=E_q[\bet]=\bs{0}$ (\refeq{eq:postq}) which was described earlier. As a result, the optimal $(\bmt,\bmz)$ can be computed {\em beforehand} and the rest of the VB-EM steps would involve only iterative updates of $q$ (i.e. $\bs{C}_{\theta \theta}, \bs{C}_{\theta y}, \bs{C}_{yy}, \tau_z$) and $\bs{W}$.

\item The optimization of $\mathcal{F}_W$ with regards to the orthogonal matrix $\bs{W}$ requires appropriate nonlinear constrained optimization tools. A highly efficient such tool is discussed in  \ref{app:w}. We reiterate that this step does not require any further calls to the forward solver.

\item We also  point out that  the proposed algorithm inherits all the favorable traits of Expectation-Maximization algorithms as  discussed in \cite{neal_view_1998}. As explained therein it suffices that the updates for $q$ (VB-E-step) or $\bs{R}$ (VB-M-step) lead to an improvement of the variational bound $\mathcal{F}$ rather than being the locally optimally values. This would for example   allow for only partial updates of $q$ (e.g. updating $\bs{C}_{\theta \theta}$ only every few iterations) or incremental improvements of $\bs{W}$ that simply lead to an increase in $\mathcal{F}_W$ without finding the local maximum. 
Such strategies could expedite significantly the computations involved.

\item  
Finally we note that implicit to the aforementioned derivations is the assumption of a unimodal density on the latent variables and as a result a unique, global maximum for $\bz$ (\refeq{eq:postz}). This assumption can be relaxed by employing a mixture of Gaussians (e.g. \cite{choudrey_variational_2003}) that will enable the approximation of highly non-Gaussian and potentially multi-modal $p_{aux}$ which in turn can reveal multiple {\em local maxima} of the expected utility $V(\bz)$. Such approximations could also be combined with the employment of different basis sets $\bs{W}$ for  each of the mixture component i.e. different sensitive/sloppy directions for each local optimum. We defer further discussions along these lines to future work.
\ei

\subsection{Validation - Assessing the accuracy of approximations}
\label{sec:val}

Thus far  we have employed the  variational lower bound in order to identify the optimal dimensionality reduction and to infer the latent variables that approximate $p_{aux}(\bz)$  (\refeq{eq:postz}) and  the expected utility $V(\bz)$ (\refeq{eq:Vapprox}). 
The goal in this section is to propose quantitative indicators that assess  the accuracy of the VB approximation.
To that end we consider the Kullback-Leibler divergence $KL(q(\by,\bez, \bet)|| p_{aux}(\by,\bez,\bet | \bs{R}))$ (\refeq{eq:kl}) that motivated the VB-EM scheme discussed.
In particular:
\be
\begin{array}{ll}
KL(q(\by,\bez, \bet)|| p_{aux}(\by,\bez,\bet | \bs{R})) & = -E_q\left[ \log \frac{ p_{aux}(\by,\bez,\bet | \bs{R})}{q(\by,\bez, \bet)} \right] \\
& = - E_q\left[\frac{ p_{aux}(\by,\bez,\bet, \bs{R})}{p_{aux}(\bs{R}) q(\by,\bez, \bet)}  \right] \\
  & = \log p_{aux}(\bs{R}) -E_q\left[\frac{ p_{aux}(\by,\bez,\bet, \bs{R})}{p_{aux}(\bs{R}) q(\by,\bez, \bet)}  \right]
  \end{array}
  \label{eq:kl2}
  \ee
  where $ p_{aux}(\by,\bez,\bet, \bs{R})$  is given in \refeq{eq:auxjoint} and $\log p_{aux}(\bs{R})$ in \refeq{eq:auxR}. We propose estimating both terms in \refeq{eq:kl2} using Importance Sampling with $q(\by,\bez, \bet)$ as the Importance Sampling density \cite{beal_variational_2003}.
  If we denote by:
  \be
w(\by,\bez, \bet)=\frac{ p_{aux}(\by,\bez,\bet, \bs{R}) }{q(\by,\bez, \bet)}.
\ee
the (un-normalized) importance weights, then by drawing samples $\{ \by^{(m)},\bez^{(m)}, \bet^{(m)}\}_{m=1}^M$  from $q(\by,\bez, \bet)$ we obtain that:
\be
\log <w>=\log \left( \frac{1}{M} \sum_{m=1}^M w(\by^{(m)},\bez^{(m)}, \bet^{(m)}) \right) \longrightarrow \log p_{aux}(\bs{R})
\ee
and:
\be
<\log w>=\frac{1}{M} \sum_{m=1}^M \log w(\by^{(m)},\bez^{(m)}, \bet^{(m)}) \longrightarrow E_q\left[\log \frac{ p_{aux}(\by,\bez,\bet, \bs{R})}{q(\by,\bez, \bet)}  \right]
\ee
In summary, by employing Importance Sampling we can estimate:
\be
KL(q(\by,\bez, \bet)|| p_{aux}(\by,\bez,\bet | \bs{R})) \approx \log <w> - <\log w >
\ee
We note that sampling from $q(\by,\bez, \bet)$ is straightforward due to its Gaussian form but the evaluation of the weights  require the computation of the actual  utility i.e. running the exact forward model. We point out however that this is done {\em solely} for the purposes of validation.
Given that the $KL$-divergence is not bounded from above and in order to compare it when considering various values of $d_y$ i.e. the dimension of the reduced coordinates $y$, we propose normalizing it with the entropy $H(q)$ of  the multivariate Gaussian  $q(\by,\bez, \bet)$ which can be exactly computed as:
\be
H(q)=-\frac{d_{\theta}+d_y}{2}\log 2\pi -\frac{1}{2}\log \left| \begin{array}{cc} \bs{C}_{\theta \theta}^{opt} & \bs{C}_{\theta y}^{opt} \\ sym. & \bs{C}_{yy}^{opt} \end{array}  \right|-\frac{d_z-d_y}{2}\log \frac{2\pi}{\tau_z^{opt}} 
\ee
In the examples that follow we report therefore the following {\em normalized} KL-divergence:
\be
nKL=\frac{ KL(q(\by,\bez, \bet)|| p_{aux}(\by,\bez,\bet | \bs{R})) }{H(q)}
\label{eq:nkl}
\ee
In all the expressions above we use $\bs{R}^{opt}$ as found at the last iteration of the VB-EM scheme.

\section{Numerical Illustrations}
\label{sec:numericalexamples}
In this section we discuss the numerical results obtained in the analysis of two  examples. In both cases the forward model consisted of an elliptic PDE. The discretization of the forward  problem for the computation of outputs $\bs{u}$ as well as of the  adjoint problem for the computation of the derivatives $\bs{G}_{\theta}, \bs{G}_z$ was performed using standard finite element tools.  Both problems involved a very high number of random variables $d_{\theta}$ arising from the discretization of a random field with small correlation length (in relation to the problem domain). Especially the second example involved a very large number of design variables $d_z$. A summary of the basic dimensions/quantities is contained in Table \ref{tab:gen}.

With regards to the regularization (prior) terms, in both problems we employed  $\tau_{y0}^{-1}=10^4$ (\refeq{eq:priory}) and $\epsilon^2=10^{-10}$ (\refeq{eq:priorez}).
Details about the $p_{\mu_z}(\bmz)$ are given for each example separately. With regards to the VB-EM scheme employed we note that at each iteration,  $100$ $\bs{W}-$updates were performed according to the equations detailed in \ref{app:w}.

\begin{table}
 \begin{tabular}{c|c|c|c|c}
   & Random Variables $\bt$ & Design Variables $\bz$ & $\tau_Q^{-1}$ & $n=dim(\bs{u}_{target})$\\ 
   \hline \hline 
   Num. Illustration 1 & $d_{\theta}=1600$ & $d_z=21$ & 0.01 & 11 \\
   \hline
   Num. Illustration 2 & $d_{\theta}=3536$ & $d_z=3536$  & $5\times 10^{-6}$ & 8 \\
 \end{tabular}
 \caption{Basic quantities/dimensions}
\label{tab:gen}
\end{table}

\subsection{Numerical Illustration 1}
\label{sec:ni1}

The goal of this problem is to optimally select the input to a random, heterogeneous medium so as to maximize an expected utility related to the response. In particular we consider the rectangular domain $\Omega = [-1,1] \times [0,1]$ of
 Figure  \ref{fig:ex1} and the steady-state heat diffusion with a governing PDE:
\be
\nabla \cdot \big( -\lambda({\boldsymbol{x}}) \nabla u(\boldsymbol{x}) \big) = 0  , \quad \bs{x} \in int(\Omega)
\label{eq:heat}
\ee
The boundary conditions are $u=0$ on $\Gamma_D$, $ -\lambda(\boldsymbol{x})\frac{\partial T(\boldsymbol{x})}{\partial n} = 0$ on $\Gamma_{N}$. The design variables $\bz$ parametrize the flux on the left hand boundary.

\begin{figure}[!ht]
 \psfrag{qn}{$\Gamma_{N}$}
 \psfrag{t0}{$\Gamma_D$}
 \psfrag{d}{$\bz$}
 \psfrag{target}{$\bs{u}_{target}$}
 \psfrag{2}{$2$}
 \psfrag{1}{$1$}
 \caption{Problem Configuration for Numerical Illustration 1 }
 \label{fig:ex1}
\end{figure}

The uncertainties $\bt$ parametrize the conductivity field $\lambda(\bs{x})$.
In particular we consider a statistically-homogeneous, log-normally-distributed  random field  with mean $1$ and coefficient of variation $0.50$.
This is defined through a transformation of a statistically-homogeneous  Gaussian field $\lambda_g(\bs{x})$ as:
\be
\lambda(\boldsymbol{x}) = e^{\lambda_g(\boldsymbol{x})}
\label{eq:map2}
\ee
The following  autocovariance $C_g(\Delta x_1, \Delta x_2)$ for   $\lambda_g(\bs{x})$  is employed:
\be
C_g(\Delta x_1, \Delta x_2)=\sigma_g^2 \exp \{-\frac{ \sqrt{\Delta x_1^2+\Delta x_2^2}}{x_0} \}, \quad \sigma_g^2=0.223
\label{eq:autocov}
\ee
where a correlation length of $x_0=0.1$ is used. We note that the correlation length is small in relation to the dimensions of the problem domain and as a result  a large number of random variables $\bt$ are required. In particular we discretize the problem domain into $1600$ triangular,  finite elements \footnote{we consider a $40 \times 20$ regular grid and each rectangle is divided along its diagonal into two triangles}  and model with $\bt$ the value of $\lambda_g(\bs{x})$ at the centroid of each element.
This gives rise to $d_{\theta}=1600$ and a $p_{\theta}$ (\refeq{eq:linp}) with mean $\bs{\mu}_{\theta 0}=-0.112$, variance $\sigma_g^2=0.223$ and covariance matrix $\bs{C}_{\theta 0}$ obtained from  \refeq{eq:autocov}. Sample realizations of the conductivity field $\lambda(\bs{x})$ are depicted in Figure \ref{fig:cfield} for illustrative purposes.

We employ a design variable $\bz$ for each node along the left-hand boundary of the problem domain (Figure \ref{fig:ex1}) resulting in $d_z=21$ design variables.
Finally, with regards to the utility function $U$, we use  temperatures  along $x_1=0, x_2 \in [-.25,0.75]$ (red line in Figure \ref{fig:ex1}) and in particular at $11$ equidistant points with $x_{2,k}=0.25+0.05 (k-1), k=1 \div 11$. The target temperature vector $\bs{u}_{target}$ is set to:
\be
 \bs{u}_{target,k}=20-40| x_{2,k}-0.5| 
\ee                         
 and $\tau_Q^{-1}=0.01$ (\refeq{eq:aux5}).
We finally note that a vague Gaussian regularization/prior was employed for $\bmz$ such that $p_{\mu_z}(\bmz) \equiv \mathcal{N}(\bs{0}, \bs{C}_{z0}= 10^{10} \bs{I})$.

\begin{figure}
 \begin{minipage}[b]{.5\linewidth}
\centering%
\subcaption{sample 1}
\end{minipage}%
\hfill%
\begin{minipage}[b]{.5\linewidth}
\centering%
\subcaption{sample 2}
\end{minipage}
\\ 
 \begin{minipage}[b]{.5\linewidth}
\centering%
\subcaption{sample 3}
\end{minipage}%
\hfill%
\begin{minipage}[b]{.5\linewidth}
\centering%
\subcaption{sample 4}
\end{minipage}
\caption{Sample realizations of the conductivity (example 1) - Young's modulus (example 2) field $\lambda(x)$ as  prescribed in Equation (\ref{eq:map2})}
\label{fig:cfield}
\end{figure}

Figure \ref{fig:ex1muz} depicts the computed $\bmz$ as a function of the number of iterations (\ref{app:mu}). As it can be seen, convergence is attained with as few as $20$ forward calls. We re-emphasize that these are the only forward solutions required for the computation of the outputs and their derivatives. We note that while the linearization in \refeq{eq:linu}  with respect to $\bz$ is exact, this is not the case with regards to the random variables $\bt$. This is due primarily to the nonlinear dependence of the response on the conductivity field $\lambda(\bs{x})$. 

\begin{figure}[!t]
 \centering
 \psfrag{y}{$x_2$}
 \psfrag{muz}{$\bmz$}
\caption{Computed $\bmz$ (see  \ref{app:mu}) as a function of the iteration number. In example 1 this expresses the flux on the left boundary $x_1=0,~x_2 \in [0,1]$. Each iteration involves a forward call for the computation of the output $\bs{u}$ and its derivatives.}
 \label{fig:ex1muz}
\end{figure}

The evolution of the  the variational lower-bound $\mathcal{F}$ (\refeq{eq:Fopt}) with regards to the  iterations alternating between $q$ and $W$ updates is shown in  Figure  \ref{fig:Fvarex1}. We note that these iterations {\em do not entail any additional forward calls}.
Figure \ref{fig:ex1s2j} depicts the evolution of the identified $\sigma_j^2$ per VB-EM iteration where as it is clearly seen, there exist   3 ``stiff''  generalized eigenvectors with small values for the corresponding generalized eigenvalues. One also notes that the variances top-off at the prior value $\tau_{y0}^{-1}=10^{4}$.
These 3 most sensitive  generalized eigenvectors  $\bs{\hat{W}}$ (\refeq{eq:what}) and the associated variances are shown in Figure \ref{fig:ex1W}. 
The numbers in parentheses were the computed variances when the calculation was repeated for exactly the same problem but by assuming a coefficient of variation of $0.71=\sqrt{0.5}$ (instead of $0.50$) for the conductivity field $\lambda(\bs{x})$. The most sensitive eigenvectors were identical (and therefore not plotted) but, as expected, their sensitivity is reduced or equivalently the corresponding variances were larger. Figure \ref{fig:ex1compare} compares the $\bmz$ computed for these two cases where one notes that while the shape is the same the amplitude/range is different. 

 Figure \ref{fig:ex1altdes} depicts  sample designs drawn from $q(\bmz+\bs{W}y)$ (which approximates the expected utility $V(\bmz+\bs{W}y))$ corresponding to different (relative) levels of the the expected utility.
While in the approximation advocated $\bmz$ represents the optimal design for which $V(\bz)$ attains its (locally) maximum value, by considering expected utility values $V(\bz)$ less than the optimal we can identify an infinity of alternative designs but also assess the sensitivity  of the solution.

Finally in Table \ref{tab:ex1} we record  the normalized KL-divergence as discussed in Section \ref{sec:val} and note that this decays for increasing $d_y$ to relatively small values indicating a good quality in the approximation found.

\begin{figure}
\begin{minipage}[b]{.5\linewidth}
\centering%
 \psfrag{iteration}{iteration}
 \psfrag{Fvar}{$\mathcal{F}$}
\subcaption{Evolution of $\mathcal{F}$ (\refeq{eq:Fopt}). }
\label{fig:Fvarex1}
\end{minipage}%
\hfill%
\begin{minipage}[b]{.5\linewidth}
\centering%
\psfrag{s2}{$\sigma_j^2$}
\psfrag{n}{$j$}
\subcaption{Evolution of $\sigma_j^2$}
\label{fig:ex1s2j}
\end{minipage}
\caption{VB-EM Each iteration corresponds to one $q$ (Equations (\ref{eq:optq}), (\ref{eq:opttau})) and one $\bs{W}$ (\refeq{eq:objW}) update}
\end{figure}

%

\begin{figure}
\psfrag{y}{$x_2$}
\begin{minipage}[b]{.31\linewidth}
\centering%
\psfrag{w}{$\bs{\hat{w}}_1$}
\subcaption{$\sigma_1^2=4.0 \times 10^{-2}$ \\ ($5.9 \times 10^{-2})$}
\end{minipage}%
\hfill%
\begin{minipage}[b]{.31\linewidth}
\centering%
\psfrag{w}{$\bs{\hat{w}}_2$}
\subcaption{$\sigma_2^2=1.5 \times 10^3$  \\($1.6 \times 10^3$)}
\end{minipage}
\hfill%
\begin{minipage}[b]{.31\linewidth}
\centering%
\psfrag{w}{$\bs{\hat{w}}_3$}
\subcaption{$\sigma_3^2=6.6 \times 10^3$ \\($6.8 \times 10^3$)}
\end{minipage}
\caption{First three most sensitive eigenvectors $\{\bs{\hat{w}}_j\}_{j=1}^3$ (\refeq{eq:what}) and associated variances $\sigma_j^2$. We note that $\sigma_3^2/\sigma_1^2 = \mathcal{O}(10^5)$}
\label{fig:ex1W}
\end{figure}

\begin{figure}
 \centering
  \psfrag{y}{$x_2$}
 \psfrag{muz}{$\bmz$}
 \caption{Comparison of $\bmz$ computed when the conductivity field $\lambda(\bs{x})$ has a coefficient of variation (cov) of $0.50$ and $0.71$. } 
 \label{fig:ex1compare}
\end{figure}

%

\begin{figure}
\psfrag{y}{$x_2$}
 \psfrag{z}{$\bz$}
 \begin{minipage}[b]{.5\linewidth}
\centering%
\subcaption{$\frac{V(\bz)}{V(\bmz)}=0.95$}
\end{minipage}%
\hfill%
\begin{minipage}[b]{.5\linewidth}
\centering%
\subcaption{$\frac{V(\bz)}{V(\bmz)}=0.75$}
\end{minipage}
\\ 
 \begin{minipage}[b]{.5\linewidth}
\centering%
 \vspace{.75cm}
\subcaption{$\frac{V(\bz)}{V(\bmz)}=0.50$}
\end{minipage}%
\hfill%
\begin{minipage}[b]{.5\linewidth}
\centering%
\subcaption{$\frac{V(\bz)}{V(\bmz)}=0.25$}
\end{minipage}
\caption{Alternative designs $z$ at various levels of expected utility $\frac{V(\bz)}{V(\bmz)}$ as compared to the optimal $\bmz$}
\label{fig:ex1altdes}
\end{figure}

%

\begin{table}
 \centering \begin{tabular}{c|c|c}
 $d_y$ & nKL (\refeq{eq:nkl}) & \\
 \hline \hline
 1 & $1.5 \times 10^{-1}$ & \\
 2 & $1.2 \times 10^{-1}$ &  \\
 5 &  $4.7 \times 10^{-2}$ & \\
 10 & $2.5 \times 10^{-2}$  & \\
 20 & $9.8 \times 10^{-3}$&  \\
 \end{tabular}
 \caption{Normalized KL-divergence from \refeq{eq:nkl} for example 1}
\label{tab:ex1}
\end{table}

\subsection{Numerical Illustration 2:  {\em Stochastic Topology Optimization}}
\label{sec:topopt}

The vast majority of studies in the context of stochastic topology optimization consider uncertainties in the loads (i.e. input) of linear systems \cite{guest_structural_2008}. This allows one to find closed-form expressions for the  random response  and perform the integrations needed much more easily.
Recently notable efforts have been made towards addressing the significantly more complicated problem involving geometric and/or material uncertainties \cite{schevenels_robust_2011}. 
Some of the proposed solution strategies  employed perturbations techniques \cite{jalalpour_optimal_2011,lazarov_topology_2012-3} whose performance decays as the random variability around the mean and/or the number of random variables increases .
Other attempts have made use of intrusive \cite{tootkaboni_topology_2012} and non-intrusive \cite{kim_efficient_2006,lazarov_topology_2012} versions of (generalized) Polynomial Chaos (gPC) in order to address the stochastic components. 

We consider the two-dimensional  domain $\Omega=[0,1.6] \times [0,1]$ in Figure \ref{fig:ex2conf}. The goal is to identify where the material of interest should be placed in order to achieve the objectives (subject to appropriate constraints) to be discussed.
We can therefore partition $\Omega$ into $\Omega_1$ which contains all points where material is placed and $\Omega_0=\Omega \ \Omega_1$ which corresponds to the points without any material (void).    
The governing differential is that of elastostatics:
\be
\begin{array}{l}
\nabla \cdot  \left( \bs{D}(\bs{x}) \bs{\epsilon}(\bs{u}(\bs{x})) \right)= \bs{0}, \quad \bs{x} \in int(\Omega) \\
 \bs{\epsilon}(\bs{u}(\bs{x}))=\left[ \begin{array}{c} \frac{\pa u_1}{\pa x_1} \\ \frac{\pa u_2}{\pa x_2} \\ \frac{\pa u_1}{\pa x_2}+\frac{\pa u_2}{\pa x_1}
 \end{array} \right]
\end{array}
\label{eq:ex2gov}
\ee
where $\bs{u}(\bs{x})=\left[ \begin{array}{c}
                              u_1(x_1,x_2) \\u_2(x_1,x_2)
                             \end{array} \right]$ is the displacement field,   
$\bs{D}$ is the (plane-stress) elasticity matrix\footnote{$\nu=0.3$ (constant) in this study} i.e. $\bs{D}(\bs{x})=\frac{E(\bs{x})}{1-\nu^2} \left[\begin{array}{ccc} 1 & \nu & 0 \\ \nu & 1 & 0 \\ 0 & 0 & 1-\nu \end{array} \right]$  and $\bs{E}(\bs{x})$ is the Young's modulus. Its spatial variation can be modeled as:
\be
E(\bs{x})=E_{min}+1_{\Omega_1}(\bs{x}) (\lambda(\bs{x})-E_{min}), \quad 1_{\Omega_1}(\bs{x})=\left\{ \begin{array}{ll}
                  0 & \textrm{if $\bs{x}\in \Omega_0$} \\
                   1 &   \textrm{if $\bs{x}\in \Omega_1$} 
                  \end{array} \right.
                  \label{eq:bin}
\ee
The value of $E_{min}=10^{-10}$ (instead of $0$) is used to avoid numerical issues in the solution of the governing equations. 
With regards to boundary conditions it is assumed that $\bs{u}=\bs{0}$ along $\Gamma_D$ and traction-free along $\Gamma_N$ (Figure \ref{fig:ex2conf}) with the exception of a point force $P=10^{-3}$ at $(x_1=1.6, x_2=0)$.

In deterministic formulations, $\lambda(\bs{x})$ is assumed constant. In the context of the analysis pursued in this study we are interested in exploring the case where $\lambda(\bs{x})$  not only varies spatially but also exhibits {\em stochastic variability} i.e. $\lambda(\bs{x})$  is a random field.
The model adopted for $\lambda(\bs{x})$ is identical to that in Example 1 which we repeat here for completeness. In particular we define $\lambda(\bs{x})$ through a transformation of a statistically-homogeneous,  Gaussian random field $\lambda_g(\boldsymbol{x})$ as in \refeq{eq:map2}. The latter has  a mean (constant) $\mu_g=-0.112$ and autocovariance $C_g(\Delta x_1, \Delta x_2)$  as  prescribed in \refeq{eq:autocov} with a correlation length $x_0=0.1$  and a variance  $\sigma_g^2=0.223$. This gives rise to a log-normally distributed $\lambda(\bs{x})$ with mean $1$ and coefficient of variation $0.50$.

The problem domain $\Omega$ is discretized using a regular mesh of $3536$ triangular elements  \footnote{we consider a $52 \times 34$ regular grid and each rectangle is divided along its diagonal into two triangles}. The vector of random variables $\bt$ represents the values of $\lambda_g(\bs{x})$ at the centroid of each element.
This gives rise to $d_{\theta}=3536$ and a $p_{\theta}$ (\refeq{eq:linp}) with mean $\bs{\mu}_{\theta 0}=-0.112$, variance $\sigma_g^2=0.223$ and covariance matrix $\bs{C}_{\theta 0}$ obtained from  \refeq{eq:autocov}. 
We note that, as in Example 1, a small correlation length is selected giving rise to  a large number of random variables $\bt$.

Normally  the design variables $\bz$ should be binary and discretize the indicator function $1_{\Omega_1}(\bs{x})$ in \refeq{eq:bin} \footnote{We note that in deterministic formulations level-set-based representation have also been adopted e.g. \cite{conti_shape_2009,chen_new_2011}}.
As in deterministic topology optimization schemes \cite{bendsoe_topology_2003} and in order to be able to compute meaningful derivatives with respect to the design variables we adopt a relaxation of the problem. In order to represent the variations of the elastic modulus $E(\bs{x})$ (\refeq{eq:bin}), we employ the sigmoid function to transform a real-valued field $z(\bs{x})$  as follows: 
\be
E(\bs{x})=E_{min}+\frac{1}{1+e^{-z(\bs{x})}}  (\lambda(\bs{x})-E_{min})
\label{eq:bin1}
\ee
While the sigmoid function ensures that $E(\bs{x}) \in [E_{min}, \lambda(\bs{x})]$ as in \refeq{eq:bin} it does not necessarily yield a {\em hard} partitioning ($0-1$) of $\Omega$ as required in such problems. To achieve this i.e. to promote solutions where $z(\bs{x}) \to -\infty$ (i.e. $E(\bs{x}) \to  E_{min}$) or  $z(\bs{x}) \to +\infty$ (i.e. $E(\bs{x}) \to \lambda(\bs{x}) $) we adopt an appropriate {\em hierarchical prior/regularization $p_{z}(\bz)$} that is discussed in detail in  \ref{app:topopt}.
Naturally the vector of design variables $\bz$ represents the values of the field $z(\bs{x})$ at the centroid of each finite element (as we did for the random variables $\bt$) resulting in 
{\em $d_{z}=3536$ design variables} (Table \ref{tab:gen}).

More importantly though the problem formulation is only meaningful with the introduction of   a   constraint on the volume of material that should be used i.e. the volume fraction  $VF=\frac{ area(\Omega_1)}{area(\Omega)}$. This in turn implies an {\em equality constraint} for the design variables $\bz$ which  can be written as:
\be
c(\bz)= \frac{1}{d_z} \sum_{j=1}^{d_z} \frac{1}{1+e^{-z_j}} -VF=0,
\label{eq:constr}
\ee
where $VF$ is the targeted volume fraction \footnote{In the example considered, the area of each finite element is the same. If this does not hold, the constraint has to be adjusted appropriately without loss of generality}.
In order to account for this nonlinear constraint in the proposed framework where the design variables $\bz$ are treated as random variables, we propose expanding the target, auxiliary $p_{aux}(\bt,\bz)$ (\refeq{eq:aux4}) as follows:
\be
p_{aux}(\bt, \bz) \propto  e^{-\frac{c^2(\bz)}{2 \epsilon_c^2} } U(\bt,\bz) p_{\theta}(\bt) ~p_z(\bz)
 \label{eq:auxconstr}
 \ee
 Clearly this represents a {\em soft, probabilistic} enforcement of the aforementioned constraint where for small  $\epsilon_c^2$, the target density $p_{aux}$, and therefore the associated $\bz$,  are contained in the vicinity of the manifold implied by \refeq{eq:constr}.
The additional term in $p_{aux}$ in \refeq{eq:auxconstr} partially alters  the associated update equations of the VB-EM  scheme previously presented. We discuss these in detail in  \ref{app:topopt} as well. In the examples presented the value $\epsilon_c^2=10^{-10}$ was used. 

For the complete definition of the problem, we note that the target response vector $\bs{u}_{target}$ consisted of the {\em vertical displacement} $u_2$ at $8$ points along the bottom boundary i.e. with $x_2=0$ and $x_1=0.2~ k, ~k=1 \div 8$ such that:
\be
u_{target,k}= 6.25 \times 10^{-3} ~k
\ee
and $\tau_Q^{-1}=5 \times 10^{-6}$ (\refeq{eq:aux5}). For comparison purposes, the deterministic problem was solved  for $VF=0.4$. To that end, the exact same algorithmic scheme for finding $\bmt, \bmz$ was employed (\ref{app:topopt}) by assuming that the variance of the random variables $\bt$ was zero and their mean exactly the same as detailed above. The resulting  $\bmz$ which is shown in Figure \ref{fig:detsol} was obtained after (approximately) $50$ iterations and exhibits two diagonal ribs that are obviously critical in stiffening  the  system. As it is easily understood, the objective function is not (in general) concave and multiple local maxima could exist.

\begin{figure}
\psfrag{s2}{$\sigma_j^2$}
\psfrag{n}{$j$}
 \begin{minipage}[b]{.5\linewidth}
\centering%
\psfrag{gd}{$\Gamma_D$}
\psfrag{gn}{$\Gamma_N$}
\psfrag{utarget}{$\bs{u}_{target}$}
\psfrag{P}{$P$}
\psfrag{omega}{$\Omega$}
\subcaption{Problem configuration}
  \label{fig:ex2conf}
\end{minipage}%
\hfill%
\begin{minipage}[b]{.5\linewidth}
\centering%
\subcaption{Deterministic solution for $VF=0.4$ obtained by setting $Var[\bt]=0$}
\label{fig:detsol}
\end{minipage}
\caption{Problem domain, boundary conditions and deterministic solution }
\end{figure}

Figure \ref{fig:ex2meanz} depicts the estimated $\bmz$ for the {\em stochastic problem} and for two volume fractions considered i.e.  $VF=0.4$ and $VF=0.2$. The first was obtained with $35$ forward calls whereas the second with $54$. As compared to the deterministic solution in Figure \ref{fig:detsol} with the two diagonal stiffening ribs, one notes that in  Figure \ref{fig:bmz04} only one is present. This could be attributed to  a different local maximum  or it could be the result of the random variability in the properties of the material. 
\begin{figure}
\psfrag{y}{$x_2$}
 \psfrag{z}{$\bz$}
 \begin{minipage}[b]{.5\linewidth}
\centering%
\subcaption{$VF=0.4$}
 \label{fig:bmz04}
\end{minipage}%
\hfill%
\begin{minipage}[b]{.5\linewidth}
\centering%
\subcaption{$VF=0.2$}
\end{minipage}
\caption{Computed $\bmz$ (see  \ref{app:mu} and \ref{app:topopt}) as a function of the iteration number. Each iteration involves a forward call for the computation of the output $\bs{u}$ and its derivatives. For  $VF=0.4$ and $VF=0.2$ the computation required $35$  and $54$ such calls respectively.}
\label{fig:ex2meanz}
\end{figure}


\begin{figure}
\psfrag{iteration}{iteration}
\psfrag{Fvar}{$\mathcal{F}$}
 \begin{minipage}[b]{.5\linewidth}
\centering%
\subcaption{$VF=0.4$}
\end{minipage}%
\hfill%
\begin{minipage}[b]{.5\linewidth}
\centering%
\subcaption{$VF=0.2$}
\end{minipage}
\caption{Evolution of $\mathcal{F}$ (\refeq{eq:Fopt}). Each iteration corresponds to one $q$ (Equations (\ref{eq:optq}), (\ref{eq:opttau})) and one $\bs{W}$ (\refeq{eq:objW}) update}
  \label{fig:ex2fvar}
\end{figure}

More importantly the algorithm proposed can identify the most sensitive directions around the local maximum. These are obtained through successive iterations  between   $q$ (Equations (\ref{eq:optq}), (\ref{eq:opttau})) and  $\bs{W}$ updates (\refeq{eq:objW}). The evolution of the  the variational lower-bound $\mathcal{F}$ (\refeq{eq:Fopt}) with regards to these iterations is depicted in  Figure  \ref{fig:ex2fvar}. We note that these iterations {\em do not entail any additional forward calls}.
Some of the generalized eigenvectors identified $\bs{\hat{W}}$ (\refeq{eq:what}) and the associated variances are shown in Figure \ref{fig:ex2wbar}. Due to the presence of the constraint, the first (most sensitive) such eigenvector is determined by the gradient of the constraint at $\bmz$ and the associated variance $\sigma_1^2$ (in parentheses, Figure \ref{fig:ex2wbar})  by the user-specified parameters $\epsilon_c$ (\refeq{eq:auxconstr}).  
Figure \ref{fig:ex2s2j} depicts the evolution of the identified $\sigma_j$ per VB-EM iteration where as it is clearly seen the first, most sensitive generalized eigenvectors are identified in the first few iterations. One also notes that the variances top-off at the prior value $\tau_{y0}^{-1}=10^{4}$.

Figure \ref{fig:ex2wbar2} depicts the squared values  $(\bs{\hat{w}}_j)^2$ (shown in Figure \ref{fig:ex2wbar}) in a log-scale. This allows one to see how the sensitivity associated with each generalized eigenvector is spatially distributed.  
Finally Figure \ref{fig:ex2altdes02} depicts the outlines of sample designs drawn from $q(\bmz+\bs{W}y)$ (which approximates the expected utility $V(\bmz+\bs{W}y))$ corresponding to different (relative) levels of the the expected utility.
In the approximation advocated, $\bmz$ represents the optimal design for which $V(\bz)$ attains its (locally) maximum value. By considering $V(\bz)$ less than the optimal, we can identify an infinity of alternative designs but  also assess the sensitivity  of the solution.

Finally in Table \ref{tab:ex2} we record  the normalized KL-divergence as discussed in Section \ref{sec:val} and note that this decays for increasing $d_y$ to relatively small values indicating a good quality in the approximation found, particularly for $VF=0.4$.

\begin{figure}
\psfrag{s2}{$\sigma_j^2$}
\psfrag{n}{$j$}
 \begin{minipage}[b]{.5\linewidth}
\centering%
\subcaption{$VF=0.4$}
\end{minipage}%
\hfill%
\begin{minipage}[b]{.5\linewidth}
\centering%
\subcaption{$VF=0.2$}
\end{minipage}
\caption{Evolution of $\sigma_j^2$ }
  \label{fig:ex2s2j}
\end{figure}


%

\begin{figure}
\psfrag{y}{$x_2$}
 \psfrag{z}{$\bz$}
\begin{minipage}[b]{.5\linewidth}
\centering%
\subcaption{($\sigma_1^2=7.31 \times 10^{-1}$)} 
\subcaption{$\sigma_2^2=1.25 \times 10^{2}$}
\subcaption{$\sigma_5^2=2.78 \times 10^{3}$}
\subcaption{$\sigma_7^2=1.36 \times 10^{4}$}
\end{minipage}%
\hfill%
%
\begin{minipage}[b]{.5\linewidth}
\centering%
\subcaption{($\sigma_1^2=3.24 \times 10^{0}$)} 
\subcaption{$\sigma_2^2=1.93 \times 10^{2}$}
\subcaption{$\sigma_5^2=1.91 \times 10^{3}$}
\subcaption{$\sigma_9^2=1.99 \times 10^{4}$}
\end{minipage}
\caption{Generalized eigenvectors $\bs{\hat{w}}_j$ for $VF=0.4$ (left column) and $VF=0.2$ (right column)}
  \label{fig:ex2wbar}
\end{figure}

\begin{figure}
\psfrag{y}{$x_2$}
 \psfrag{z}{$\bz$}
\begin{minipage}[b]{.5\linewidth}
\centering%
\subcaption{$(\bs{\hat{w}}_1)^2$} 
\subcaption{$(\bs{\hat{w}}_2)^2$}
\subcaption{$(\bs{\hat{w}}_5)^2$}
\subcaption{$(\bs{\hat{w}}_7)^2$}
\end{minipage}%
\hfill%
%
\begin{minipage}[b]{.5\linewidth}
\centering%
\subcaption{$(\bs{\hat{w}}_1)^2$} 
\subcaption{$(\bs{\hat{w}}_2)^2$}
\subcaption{$(\bs{\hat{w}}_5)^2$}
\subcaption{$(\bs{\hat{w}}_9)^2$}
\end{minipage}
\caption{The squares of the entries of each of the generalized eigenvectors $\bs{\hat{w}}_j$ (log scale) for $VF=0.4$ (left column) and $VF=0.2$ (right column)}
  \label{fig:ex2wbar2}
\end{figure}

\begin{figure}
\psfrag{y}{$x_2$}
 \psfrag{z}{$\bz$}
 \begin{minipage}[b]{.33\linewidth}
\centering%
\subcaption{$\frac{V(\bz)}{V(\bmz)}=0.75$}
\end{minipage}%
\hfill%
\begin{minipage}[b]{.33\linewidth}
\centering%
\subcaption{$\frac{V(\bz)}{V(\bmz)}=0.50$}
\end{minipage}
\begin{minipage}[b]{.33\linewidth}
\centering%
\subcaption{$\frac{V(\bz)}{V(\bmz)}=0.25$}
\end{minipage}
\caption{Outline of alternative designs $z$ at various levels of expected utility $\frac{V(\bz)}{V(\bmz)}$ as compared to the optimal $\bmz$ ($VF=0.4$)}
\label{fig:ex2altdes04}
\end{figure}

\begin{figure}
\psfrag{y}{$x_2$}
 \psfrag{z}{$\bz$}
 \begin{minipage}[b]{.33\linewidth}
\centering%
\subcaption{$\frac{V(\bz)}{V(\bmz)}=0.75$}
\end{minipage}%
\hfill%
\begin{minipage}[b]{.33\linewidth}
\centering%
\subcaption{$\frac{V(\bz)}{V(\bmz)}=0.50$}
\end{minipage}
\begin{minipage}[b]{.33\linewidth}
\centering%
\subcaption{$\frac{V(\bz)}{V(\bmz)}=0.25$}
\end{minipage}
\caption{Outline of alternative designs $z$ at various levels of expected utility $\frac{V(\bz)}{V(\bmz)}$ as compared to the optimal $\bmz$ ($VF=0.2$)}
\label{fig:ex2altdes02}
\end{figure}

\begin{table}[!h]
\centering
\begin{tabular}{c|c|c}
  & \multicolumn{2}{c}{nKL (\refeq{eq:nkl})}  \\
  \hline
  $d_y$ & $VF=0.4$ & $VF=0.2$ \\
 \hline \hline
 5 &  $1.5 \times 10^{-2}$ & $3.4 \times 10^{-1}$ \\
 10 & $8.7 \times 10^{-3}$  & $1.9 \times 10^{-1}$\\
  15 & $3.9 \times 10^{-3}$   & $1.3 \times 10^{-1}$ \\
 20 & $6.0 \times 10^{-4}$  &  $6.8 \times 10^{-2}$  \\
 \end{tabular}
 \caption{Normalized KL-divergence from \refeq{eq:nkl} for example 2}
\label{tab:ex2}
\end{table}

\section{Conclusions}
We present a framework for solving a large class of model-based, optimization-under-uncertainty problems. 
The overarching idea is that of recasting the problem as one of probabilistic inference. This enables the uniform treatment of both random and design variables and is capable of furnishing not only a (local) maximum  (i.e. a point estimate) but also the sensitivity of the objective  to the design variables.
To achieve this objective, we propose a Variational Bayesian framework that operates on two fronts. Firstly, it attempts to compute efficiently an accurate approximation of the  joint density of interest. Secondly, it seeks  a lower-dimensional subspace with regards to the design variables $\bz$ that provides an assessment of the solution’s robustness by discovering the most sensitive directions i.e. the directions along which, variations in $\bz$ 
will cause the largest decrease in the expected utility. This is based on the same  premise 
 as the so-called  Sloppy Models whose behavior depends only on a few
stiff combinations of parameters,  with many sloppy
parameter directions largely unimportant for model behavior.
The identification of this lower-dimensional subspace, enables the analyst to compute, apart from the optimal design, an infinity of alternative designs which achieve a lower value of the expected utility.
Interestingly enough, addressing the probabilistic inference task under the Variational Bayesian perspective involves the solution of an optimization problem. To that end we propose an iterative VB-Expectation-Maximization scheme.

The aforementioned claims have been validated in the context of two numerical examples involving $\mathcal{O}(10^3)$ random and design variables. In all cases considered the cost of the computations in terms of calls to the forward model was of the order $\mathcal{O}(10 \div 10^2)$. The accuracy of the approximations provided is assessed by appropriate information-theoretic metrics. 

The framework proposed cannot currently account for the possibility of multiple local maxima, as the approximation constructed is based on unimodal Gaussian densities. Nevertheless, the formulation can be readily extended by employing  {\em mixture of Gaussians} that will enable not only  approximations for  multi-modal cases but also produce better results for unimodal, but highly non-Gaussian densities. 
We note finally the possibility of using approximate, surrogate or reduced-order models in order to expedite computations. All the algorithmic steps discussed can be readily performed by using these less-expensive forward solvers. As  long as these convey some information about the  expensive, reference forward model, then they can provide a good starting point for further computations that would require fewer expensive calls to converge.


\newpage
\appendix
\section{Maximization of $\mathcal{F}_{W}$}
\label{app:w}
As  discussed earlier, in order to update $\bs{W}$  it suffices to consider only  $\mathcal{F}_W(\bs{W})$ (\refeq{eq:objW}): 
 \be
 \begin{array}{ll}
 \mathcal{F}_{W}(\bs{W})&  = -\frac{\tau_Q}{2} \bs{W}^T \bgz^T \bgz \bs{W}: (\bs{C}_{yy}-\tau_z^{-1} \bs{I}) \\
 & - \frac{\tau_Q}{2}  2\bgt^T \bgz \bs{W} : \bs{C}_{\theta y}+\log p_{W}(\bs{W})
 \label{eq:optfw1}
 \end{array}
 \ee
 While the first part is quadratic with respect to $\bs{W}$ the difficulty arises from the orthogonality constraint $\bs{W}^T \bs{W}=\bs{I}$ which can be enforced directly or through the regularization term $p_{W}(\bs{W})$ as previously discussed.  To address this constrained optimization problem,   we employ the iterative  algorithm proposed in \cite{wen_feasible_2012} which is  highly efficient not only in terms of the number of iterations needed but also in terms of the  the cost per iteration.  It is based on the constraint-preserving Cayley transform according to which  the current $\bs{W}$ is updated to $\bs{W}'$ as follows:
 \be
 \bs{W}'=(\bs{I}+\frac{a}{2} \bs{A})^{-1} (\bs{I}-\frac{a}{2} \bs{A}) \bs{W}
 \ee
 where:
 \be
 \bs{A}=\bs{J} \bs{W}^T-\bs{W}\bs{J}^T
 \ee
 and $\bs{J}=\frac{\pa \mathcal{F}_W}{\pa \bs{W}}$.  The latter can be readily obtained from \refeq{eq:optfw1}:
 \be
 \bs{J}=-\tau_Q \bgz^T \bgz \bs{W} (\bs{C}_{yy}-\tau_z^{-1} \bs{I})-\tau_Q \bgz^T  \bgt  \bs{C}_{\theta y}
\ee
It can be shown that $\bs{W}'$ satisfies automatically the orthogonality constraint and that  and for $a=0$, $\bs{W}'$ is an ascent direction of $\mathcal{F}_W$. Several options exist for selecting the step size $a$. In the numerical illustrations we made use of the 
 Barzilai-Borwein scheme detailed in \cite{barzilai_two-point_1988} which results in a non-monotone line search algorithm.
We  note that the inversion of the $d_z\times d_z$ matrix $(\bs{I}+\frac{a}{2} \bs{A})$ can be efficiently performed  by inverting a matrix of dimension $2 d_y \times 2 d_y$ which is much smaller than $d_z$ \cite{wen_feasible_2012}. We finally re-emphasize that the updates of $\bs{W}$ require no forward calls. The updates/iterations are terminated when no further improvement to the objective $\mathcal{F}_W$ is possible.

\section{Maximization of $\mathcal{F}_{\mu}$}
\label{app:mu}
As it was previously discussed, in order to update $\bmt, \bmz$  it suffices to consider only  $\mathcal{F}_{\mu}(\bmz,\bmt)$ (\refeq{eq:optmu}): 

\be
\begin{array}{ll}
\mathcal{F}_{\mu}(\bmz,\bmt) = &-\frac{\tau_Q}{2} \left( |\bs{u}_{target}-\bs{u}(\bmt, \bmz)|^2 \right)  \\
& -\frac{1}{2}(\bmt-\bs{\mu}_{\theta 0})^T \bs{C}_{\theta 0}^{-1}(\bmt-\bs{\mu}_{\theta 0}) +\log p_{\mu_z}(\bmz)
\end{array}
\label{eq:optmu1}
\ee
This represent a nonlinear, unconstrained optimization problem that can be solved with any of the well-known algorithms \cite{nocedal_numerical_1999,boyd_convex_2004}. We present here a Gauss-Newton type algorithm that we employed and produced the results discussed in Section \ref{sec:numericalexamples}.
For clarity of the presentation we consider first the case in the first numerical illustration where the regularization/prior $p_{\mu_z}(\bmz)$ was a Gaussian $\mathcal{N}(\bs{0}, \bs{C}_{z0})$ in which case:
\be
\begin{array}{ll}
\mathcal{F}_{\mu}(\bmz,\bmt) = &-\frac{\tau_Q}{2} \left( |\bs{u}_{target}-\bs{u}(\bmt, \bmz)|^2 \right)  \\
& -\frac{1}{2}(\bmt-\bs{\mu}_{\theta 0})^T \bs{C}_{\theta 0}^{-1}(\bmt-\bs{\mu}_{\theta 0}) -\frac{1}{2} \bmz^T \bs{C}_{z0}^{-1} \bmz
\end{array}
\ee
If $(\bmz^{(t)},\bmt^{(t)})$ denote the values at iteration  $t$ and $(\bmz^{(t+1)}=\bmz^{(t)}+\Delta \bmz^{(t)},\bmt^{(t+1)}=\bmt^{(t)}+\Delta \bmt^{(t)})$, then a first-order Taylor series yields the following approximation:
\be
\begin{array}{ll}
\mathcal{F}_{\mu}(\Delta \bmz^{(t)},\Delta \bmt^{(t)}) \approx &-\frac{\tau_Q}{2} \left( |\bs{u}_{target}-\bs{u}(\bmz^{(t)}, \bmz^{(t)})-\bs{G}_{\theta,t} \Delta \bmt^{(t)}- \bs{G}_{z,t}^{(t)}  \Delta \bmz^{(t)})|^2 \right)  \\
& -\frac{1}{2}(\bmt^{(t)}+\Delta \bmt^{(t)}-\bs{\mu}_{\theta 0})^T \bs{C}_{\theta 0}^{-1}(\bmt^{(t)}+\Delta \bmt^{(t)}-\bs{\mu}_{\theta 0}) \\
& -\frac{1}{2} (\bmz^{(t)}+\Delta \bmz^{(t)})^T \bs{C}_{z0}^{-1} (\bmz^{(t)}+\Delta \bmz^{(t)})
\end{array}
\label{eq:objquad}
\ee
where $\bs{G}_{\theta,t} =\frac{\pa \bs{u}}{\pa \bt}|_{\bt=\bmt^{(t)}}$ and $\bs{G}_{z,t}=\frac{\pa \bs{u}}{\pa \bz}|_{\bz=\bmz^{(t)}}$.
Differentiating with respect to $(\Delta \bmz^{(t)},\Delta \bmt^{(t)})$ leads to the following system of coupled linear equations:
\be
\left[ \begin{array}{l}  \frac{\pa \mathcal{F}_{\mu}^{(t)} }{\pa \Delta \bmt^{(t)}}\\ 
\frac{\pa \mathcal{F}_{\mu}^{(t)} }{\pa \Delta \bmz^{(t)}} \end{array} \right]=
\left[  \begin{array}{l} \bs{0} \\ \bs{0} \end{array} \right]
\to  \bs{H}_t \left[ \begin{array}{l} \Delta \bmt^{(t)} \\ \Delta \bmz^{(t)} \end{array} \right] = \bs{h}_t 
\label{eq:updmu}
\ee
where:
\be
\bs{H}_t=\left[ \begin{array}{ll}
 \tau_Q \bs{G}_{\theta,t}^T \bs{G}_{\theta,t}+\bs{C}_{\theta 0}^{-1} & \tau_Q\bs{G}_{\theta,t}^T \bs{G}_{z,t} \\
\tau_Q\bs{G}_{z,t}^T \bs{G}_{\theta,t} &  \tau_Q \bs{G}_{z,t}^T \bs{G}_{z,t}+\bs{C}_{z0}^{-1}\\
\end{array} \right]
\ee
and:
\be
\bs{h}_t=\left[ \begin{array}{l} 
\tau_Q \bs{G}_{\theta,t}^T (\bs{u}_{target}-\bs{u}(\bmz^{(t)}, \bmz^{(t)}))-\bs{C}_{\theta 0}^{-1}(\bmt^{(t)}-\bs{\mu}_{\theta 0}) \\
\tau_Q \bs{G}_{z,t}^T (\bs{u}_{target}-\bs{u}(\bmz^{(t)}, \bmz^{(t)}))-\bs{C}_{z0}^{-1} \bmz^{(t)}
\end{array} \right]
\ee
 
We  note that at each iteration the forward solver needs to be called for the computation of the output vector $\bs{u}$ and its derivatives $\bs{G}_{\theta,z}$.  Iterations are terminated when no further improvement is possible i.e. $\frac{|\Delta \bmt^{(t)}|}{|\bmt^{(t)}|}, \frac{|\Delta \bmz^{(t)} | }{ | \bmz^{(t)}|} < (tolerance)=10^{-5}$.

\section{Regularization of $\bmz$ and update equation for Num. Illustration 2}
\label{app:topopt}
\begin{figure}

 \begin{minipage}[b]{.5\linewidth}
\centering%
\psfrag{pz}{$p_{\mu_z}(\mu_{z,j} | I_j)$}
 \includegraphics[width=.95\textwidth]{bimodal.eps}
\subcaption{Bimodal prior $p_{\mu_z}(\mu_{z,j} | I_j)$}
\label{fig:biprior}
\end{minipage}%
\hfill%
\begin{minipage}[b]{.5\linewidth}
\centering%
\psfrag{j}{$j$}
\psfrag{k}{$k$}
 \includegraphics[trim=2cm 2.5cm 2cm 2cm,clip,width=.95\textwidth]{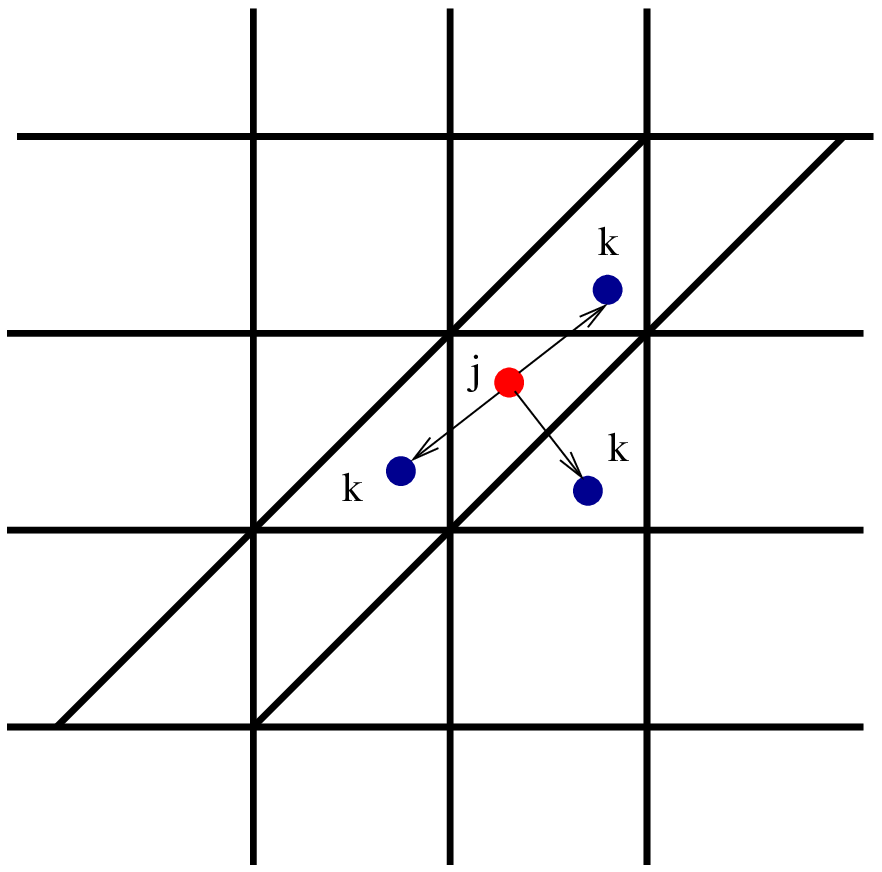}
\subcaption{Neighborhood structure}
\label{fig:neigh}
\end{minipage}
\caption{Definition of $p_{\mu_z}(\bmz)$}
\end{figure}

We discuss in this section the definition of the regularization/prior $p_{\mu_z}(\bmz)$ for numerical illustration 2 (Section \ref{sec:topopt}) and the resulting changes in the optimization scheme for $\bmz$ in \ref{app:mu}.
Given the physical interpretation of the design variables $\bmz$ as binary variables which for each pixel indicate the presence or not of material, we adopt a regularization for $\bmz$ that promotes the discovery of such solutions but also exhibits the requisite spatial correlation.
To that end we propose a hierarchical prior where in addition to $\bmz=\{ \mu_{z,j}\}_{j=1}^{3536}$ we introduce the binary hyperparameters $\bs{\phi} =\{  \phi_j=\pm 1\}_{j=1}^{3536}$ such that:
\be
p_{\mu_z}(\bmz | \bs{\phi}) = \prod_{j=1}^{3536} p(\mu_{z,j} | \phi_j)
\ee
where $p(\mu_{z,j} | \phi_j=-1)=\mathcal{N}(-m,s^2)$ and $p(\mu_{z,j} | \phi_j=+1)=\mathcal{N}(m,s^2)$. The value of $m$ was selected so that in combination with the sigmoid function (\refeq{eq:bin1}) produces solutions close to the binary images we would like to achieve:
\be
\frac{1}{1+e^{m}}=10^{-3} \approx 0, \quad \frac{1}{1+e^{-m}}=1-10^{-3} \approx 1
\ee
This yields $m=-6.9$ and the  resulting, bimodal, hierarchical prior is depicted in Figure \ref{fig:biprior}.
In order to account for the spatial dependence of neighboring $\mu_{z,j}$ we employ an auto-logistic  hyperprior on $\bs{\phi}$  of the following form \cite{besag_nearest-neighbour_1972,besag_spatial_1974}:
\be
p(\bs{\phi} | \beta)\propto e^{-\frac{\beta}{2} \sum_{j} \sum_{k \sim j} \phi_j \phi_k}
\ee
The second sum in the expression above is over all indices $k$ which correspond to sites neighboring to $j$ (neighborhood relation denoted by $\sim$). Given the triangular mesh used, we consider 3 neighbors for each site $j$ as shown in Figure \ref{fig:neigh}. 
The hyperparameter $\beta$ controls the strength of spatial correlation. At one extreme, if $\beta \to +\infty$, neighboring $\phi_j$ prefer to have different values (i.e. $-1/+1$ or $+1/-1$) as this yields a higher hyperprior value.  At the other extreme, if $\beta \to -\infty$, neighboring $\phi_j$ prefer to have the same values (i.e. $-1/-1$ or $+1/+1$). For $\beta=0$ no 
correlation is present. We note that the  aforementioned prior in $\bs{\phi}$ imbues indirectly
 spatial correlation in $\mu_{z,j}$.

In summary,  the prior $p_{\mu_z}(\bmz)$ can be found by integrating out the hyperparameters $\bs{\phi}$ and $\beta$ as:
\be
p_{\mu_z}(\bmz)=\int p_{\mu_z}(\bmz | \bs{\phi}) p(\bs{\phi} | \beta) ~d\bs{\phi} d\beta
\ee
The integration above cannot be performed analytically and for that reason we employed an Expectation-Maximization scheme \cite{dempster_maximum_1977,bishop_pattern_2007} whereby at the Expectation step a Metropolized-Gibbs scheme is  used to sample the hyperparameters $\bs{\phi}$ and $\beta$ from their conditional posterior (given the current value of $\bmz$). This does not require any forward calls and can be very efficiently performed. The samples generated can be used to estimate $\log p_{\mu_z}(\bmz)$ and its derivatives as needed for the update equations in \ref{app:mu}. 
If we denote with $<~>$ expectations with regards to the posterior samples of $\bs{\phi}$ described above and by keeping only terms that depend on $\bmz$ we obtain that:
\be
\begin{array}{ll}
 \log p_{\mu_z}(\bmz) & = -\frac{1}{2 s^2}  \sum_j < ( \mu_{z,j} -m \phi_j)^2 >  \\
  & = -\frac{1}{2 s^2} (\bmz^T \bmz-2 m \bmz^T < \bs{\phi}>+ < \bs{\phi}^T \bs{\phi}>)
\end{array}
\ee
The quadratic form of this expression implies that the only changes in the update \refeq{eq:updmu} in \ref{app:mu} will be in:
\be
\bs{H}_t=\left[ \begin{array}{ll}
 \tau_Q \bs{G}_{\theta,t}^T \bs{G}_{\theta,t}+\bs{C}_{\theta 0}^{-1} & \tau_Q\bs{G}_{\theta,t}^T \bs{G}_{z,t} \\
\tau_Q\bs{G}_{z,t}^T \bs{G}_{\theta,t} &  \tau_Q \bs{G}_{z,t}^T \bs{G}_{z,t}+\frac{1}{s^2}\bs{I}\\
\end{array} \right]
\ee
and:
\be
\bs{h}_t=\left[ \begin{array}{l} 
\tau_Q \bs{G}_{\theta,t}^T (\bs{u}_{target}-\bs{u}(\bmz^{(t)}, \bmz^{(t)}))-\bs{C}_{\theta 0}^{-1}(\bmt^{(t)}-\bs{\mu}_{\theta 0}) \\
\tau_Q \bs{G}_{z,t}^T (\bs{u}_{target}-\bs{u}(\bmz^{(t)}, \bmz^{(t)}))-\frac{1}{s^2}( \bmz^{(t)}-m < \bs{\phi}>)
\end{array} \right]
\ee

The aforementioned equations should be augmented by the equality constraint in \refeq{eq:constr}. We enforce this constraint directly on $\bmz$ so as the optimal design ($\bmz$) satisfies it. From an algorithmic point, the process adopted is similar to Sequential Quadratic Programming (SQP, \cite{nocedal_numerical_1999}) where the quadraticized objective (\refeq{eq:objquad})  at each iteration $t$ is augmented by the linearized constraint:
\be
0=c(\bmz^{(t)}+\Delta \bmz^{(t)}) \approx c(\bmz^{(t)})+\bs{f}_t^T \Delta \bmz^{(t)}
\ee
where $\bs{f}_t=\frac{\pa c}{\pa \bz}|_{ \bz=\bmz^{(t)} }$.

In order to account for the constraint in the rest of the auxiliary density $p_{aux}$, the scheme described in \refeq{eq:auxconstr} is adopted which induces  a soft/probabilistic enforcement. The term $^{-\frac{c^2(\bz)}{2 \epsilon_c^2} }$ will therefore yield an additional contribution in the variational lower-bound $\mathcal{F}$ detailed in \refeq{eq:ftot}. If we denote by $\mathcal{F}_c$ this additional term, then:
\be
\begin{array}{ll}
 \mathcal{F}_c & = -\frac{1}{2 \epsilon_c^2} E_q[c^2(\bz)] \\
 & = -\frac{1}{2 \epsilon_c^2} E_q[c^2(\bmz +\bs{W}\by+\bez)] \\
\end{array}
\label{eq:Fc}
\ee
Given the nonlinear form of $c(\bz)$, we employ another linearization around $\bmz$:
\be
\begin{array}{ll}
 c(\bmz +\bs{W}\by+\bez)  & \approx c(\bmz)+\bs{f}^T(\bs{W}\by+\bez) \\
 & = \bs{f}^T(\bs{W}\by+\bez) \quad  (\textrm{since $c(\bmz)=0$})
\end{array}
\ee
where $\bs{f}=\frac{\pa c}{\pa \bz}|_{ \bz=\bmz}$. As a result of this and the form of $q$ (\refeq{eq:postq}), \refeq{eq:Fc} becomes:
\be
\begin{array}{ll}
 \mathcal{F}_c &= -\frac{1}{2 \epsilon_c^2} \left( \bs{W}^T \bs{f~f}^T \bs{W}:\bs{C}_{yy} + \tau_{z}^{-1} \bs{f~f}^T: (\bs{I}-\bs{W}\bs{W}^T) \right)
\end{array}
\ee
which when combined with the rest of the terms in $\mathcal{F}$ in \refeq{eq:Fopt}, leads to the following changes in the update equations in the VB-EM scheme:
\bi
\item \textbf{VB-Expectation}:
\be
\left[ \begin{array}{cc} \bs{C}_{\theta \theta}^{opt} & \bs{C}_{\theta y}^{opt} \\ sym. & \bs{C}_{yy}^{opt} \end{array} \right]^{-1} =
\left[ \begin{array}{cc}  \tau_Q \bgt^T \bgt+\bs{C}_{\theta 0}^{-1} &  \tau_Q \bgt^T \bgz \bs{W} \\
    sym. & \tau_Q \bs{W}^T\bgz^T \bgz \bs{W}+\tau_{y0} \bs{I}+\frac{1}{ \epsilon_c^2}\bs{W}^T \bs{f~f}^T \bs{W}
\end{array} \right]
\ee
and:
\be
\tau_z^{opt}=\tau_{z0}+\frac{1}{d_z-d_y} (\tau_Q \bgz^T \bgz+\frac{1}{ \epsilon_c^2}\bs{f~f}^T) : (\bs{I}-\bs{W}\bs{W}^T)
\ee

\item \textbf{VB-Maximization}: 
\be
 \bs{W}^{opt}=\arg \max_{\bs{W}} \mathcal{\hat{F}}_{W}(\bs{W})
\ee
where:
\be
\begin{array}{ll}
\mathcal{F}_{W}(\bs{W}) 
  = & -(\frac{\tau_Q}{2} \bs{W}^T \bgz^T \bgz \bs{W}+\frac{1}{ 2 \epsilon_c^2}\bs{W}^T \bs{f~f}^T \bs{W}): (\bs{C}_{yy}-\tau_z^{-1} \bs{I}) \\
 & - \frac{\tau_Q}{2}  2\bgt^T \bgz \bs{W} : \bs{C}_{\theta y}+\log p_{W}(\bs{W})
\end{array}
\ee

\ei

\newpage
\bibliographystyle{elsarticle-num}
\bibliography{/home/psk/RESEARCH_DOC/my_zote_zero_library}







\end{document}